\documentclass[a4paper,nofootinbib,preprintnumbers,twocolumn,preprintnumbers,floatfix,superscriptaddress,prl,showpacs]{revtex4-1}

\usepackage{graphicx}
\usepackage{amssymb}
\usepackage{amsmath}
\usepackage{epstopdf}
\usepackage{graphics}
\usepackage[caption=false]{subfig}
\usepackage{float}
\usepackage{color}
\usepackage{hyperref}

\allowdisplaybreaks[1]

\newcommand{\bra}[1]{\langle #1|}					
\newcommand{\ket}[1]{|#1\rangle}					

\newcommand{\abs}[1]{\left| #1 \right|} 

\newcommand{\figref}[1]{Fig.~\ref{#1}}
\newcommand{\tabref}[1]{Tab.~\ref{#1}}

\newcommand{\specialcell}[2][c]{%
  \begin{tabular}[#1]{@{}c@{}}#2\end{tabular}}

\begin{document}

\title{Heralded quantum gates with integrated error detection in optical cavities}

\author{J. Borregaard }
\affiliation{The Niels Bohr Institute, University of Copenhagen, Blegdamsvej 17, DK-2100 Copenhagen \O, Denmark}
\author{P. K\'om\'ar}
\affiliation{Department of Physics, Harvard University, Cambridge, MA 02138, USA}
\author{E. M. Kessler} 
\affiliation{Department of Physics, Harvard University, Cambridge, MA 02138, USA}
\affiliation{ITAMP, Harvard-Smithsonian Center for Astrophysics, Cambridge, MA 02138, USA}
\author{A. S. S\o rensen}
\affiliation{The Niels Bohr Institute, University of Copenhagen, Blegdamsvej 17, DK-2100 Copenhagen \O, Denmark}
\author{M. D. Lukin}
\affiliation{Department of Physics, Harvard University, Cambridge, MA 02138, USA}

\date{\today}

\begin{abstract}
We propose and analyze heralded quantum gates between qubits in optical cavities. They employ an
auxiliary qubit to report if a successful gate occurred.  In this manner, the errors, which
would have corrupted a deterministic gate, are converted into a non-unity
probability of success: once successful the gate has a much higher
fidelity than a similar deterministic gate. Specifically, we describe that a
heralded , near-deterministic controlled phase gate (CZ-gate) with the conditional error arbitrarily close to zero and the success probability that approaches unity as the cooperativity
of the system, $C$, becomes large. Furthermore, we describe an extension to near-deterministic $N$-qubit Toffoli gate with a favorable error scaling. These gates can be directly employed in quantum repeater networks to facilitate near-ideal entanglement swapping, 
thus greatly speeding up the entanglement distribution.             
\end{abstract}

\pacs{03.67.Pp, 03.67.Hk, 03.67.Bg, 32.80.Qk, 42.50.Pq}

\maketitle

Exploiting quantum systems for information processing offers many potential
advantages over classical information processing like highly secure quantum networks~\cite{cirac,kimble,
duan3}
and powerful quantum computers~\cite{ladd,shor,feynman}. One of the main challenges
for the realization of functional quantum computers is to
perform gates with sufficiently high quality so that the remaining errors can be
suppressed by error correction codes, which makes the computation fault tolerant
\cite{knill2}. At the same time, applications to long distance quantum communication can be enabled by quantum repeaters, which combine
probabilistic entanglement generation over short distances with subsequent entanglement connection steps \cite{duan3}. For
these protocols, the probabilistic nature of the entanglement generation is acceptable, but it is essential that high fidelity entanglement is achieved
conditioned on a heralding measurement. Experimentally, such high fidelity
entanglement is often much easier to implement and may be realized in situations
where it is impossible to perform any quantum operations 
deterministically. Here we introduce a similar concept for gate operations and
develop the concept of  heralded quantum gates  with integrated error detection. In
the resulting gate, the infidelity, which would be present for a deterministic
gate is converted into a failure probability, which is heralded by an auxiliary
atom. Once successful, the resulting gate can have an arbitrarily small error. Such heralded gates
could facilitate fault tolerant quantum computation since detectable errors may be easier to correct than undetectable errors~\cite{grassi,ralph05,varnava}.  Alternatively it can be directly
incorporated into quantum repeater architectures for long distance quantum
communication. 

Optical cavities are ideal for conversion between the stationary gate
qubits and flying qubits (photons), which is fundamental
for quantum networks \cite{ritter, komar, acin}. Quantum gates can, in principle, also
be directly implemented in optical cavities \cite{pellizari}, but the experimental
requirements for this are very challenging due to spontaneous emission and cavity loss. The essential parameter quantifying
this is the cooperativity of the atom-cavity system, $C$. It has been argued
that directly implementing gates in optical cavities leads to a poor error scaling $1-F\propto1/\sqrt{C}$, where $F$
is the fidelity of the gate \cite{kastoryano, Anders2prl}. However, as a result of the integrated 
error detection, the heralded gates that we propose exhibit high fidelities when
successful. This enables efficient 
entanglement swapping and removes the necessity of intermediate entanglement 
purification in quantum repeaters thus increasing the distribution rate significantly. Compared 
to using other deterministic, cavity based gates, an increase in the rate of up to two 
orders of magnitude can be achieved for modest cooperativities ($<100$) and a distance of 1000 km \cite{repeater}.  

\begin{figure}
\centering
\subfloat {
\label{fig:qubita}\includegraphics[width=0.23\textwidth,height=1.55in]{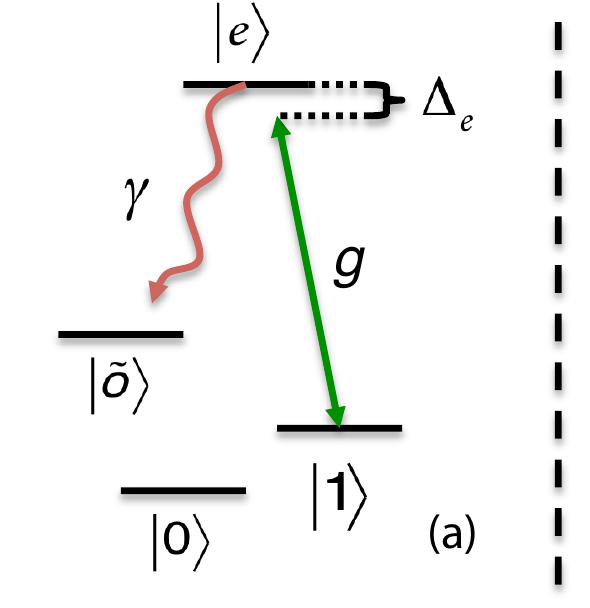}} 
\subfloat{\label{fig:catom}\includegraphics[width=0.23\textwidth,height=1.55in]{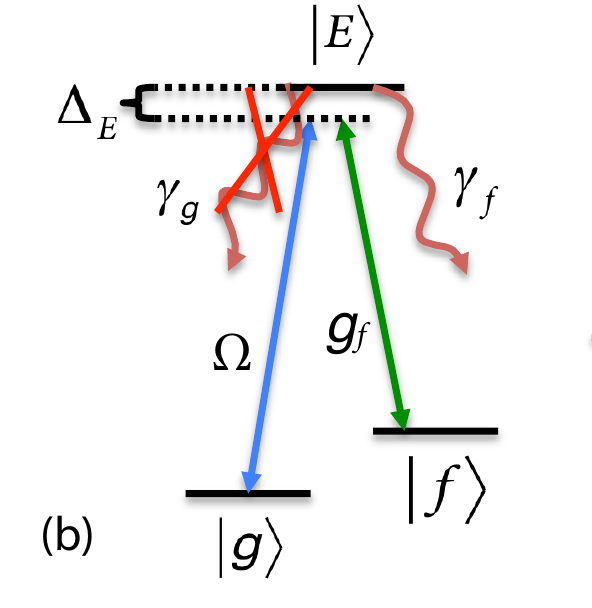}}
\caption{(Color online) (a) Level structure of the qubit atoms. Only state $\ket{1}$ couples to the cavity and we assume that the excited level decays to some level $\ket{\tilde{o}}$, possible identical to $\ket{f}$ or $\ket{0}$. (b) Level structure of the auxiliary atom and the transitions driven by the weak laser ($\Omega$) and the cavity ($g_{f}$). We assume that $\ket{E}\leftrightarrow\ket{f}$ is a closed transition, i.e. $\gamma_{g}=0$.  }
\label{fig:levels}
\end{figure}

The basic idea is to use a
heralding auxiliary atom in addition to qubit atoms in the same cavity. One of
the atomic qubit states, e.g., state $\ket{1}$ couples to the cavity mode while $\ket{0}$ is completely
uncoupled (see \figref{fig:qubita}). Such a system has previously been
considered for two-qubit gates~\cite{duan1,duan2,Anders1prl, Anders2prl,ritter2014}, multi-qubit gates~\cite{duan1,zheng} and photon routing~\cite{Tiecke}. 
If any of the qubit atoms is in state $\ket{1}$ the cavity resonance is shifted compared to
the bare cavity mode, which can be exploited to make a gate between two or more
qubits by reflecting single photons off the cavity \cite{duan1}. The efficiency
of such schemes, however, is limited by photon losses, inefficient detectors and
non-ideal single photon sources \cite{Tiecke,ritter2014}. We circumvent these
problems by introducing an auxiliary atom in the cavity to serve as both an
intra-cavity photon source and a detector. As opposed to previous heralded
gates in optical cavities, which relied on the null detection of photons leaving
the cavity \cite{pachos,beige, vitali}, the final heralding measurement on
the atom can then be performed very efficiently. 

In our approach, the auxiliary atom has
two metastable states $\ket{g},\ket{f}$, which can be coupled through an excited
state $\ket{E}$ (see \figref{fig:qubita}). We assume the
$\ket{E}\leftrightarrow\ket{f}$ transition to be energetically close to the
cavity frequency and to be a nearly closed transition,  so that we need to drive
the $\ket{g}\to \ket{E}$ transition, e.g with a two-photon process, (see below).
The gate can be understood through the phase evolution imposed on the atoms. We consider adiabatic excitation of the auxiliary control atom via Stimulated Raman Adiabatic Passage~\cite{stirap,stirap2}, driven by an external driving pulse with Rabi frequency $\Omega(t)$ and a coupling to the cavity photon $g_{f}$. In the case when all the qubit atoms are in the non-coupled states $|00..0\rangle$, an adiabatic excitation will result in a dark state $\sim g_{f} \ket{0,g} -\Omega \ket{1,f}$ with zero energy and vanishing phase. Here the number refers to the number of cavity photons.  However, the qubit states $\Psi$ with at least one of the qubit atoms in the coupled state, results in a cavity-induced shift of the state $|1,f, \Psi\rangle$, which in turn, causes an AC Stark shift and dynamical phase to be imprinted into the $|g,\Psi\rangle$ state after the driving pulse is turned off. All states but the completely uncoupled qubit state $|00...0\rangle$ will thus acquire a phase, the magnitude of which depends on the length of the driving
pulse. With an appropriate pulse length and simple single qubit rotations, we
can use this to realize a general $N$-qubit Toffoli gate or a control-phase (CZ)
gate.

Naively, the gates will be limited by errors originating from cavity decay and
spontaneous emission from the atoms, which carry away information about the
qubit state. These errors are, however, detectable since the auxiliary atom will be trapped in state $\ket{f}$ if either
a cavity excitation or an atomic excitation is lost. Conditioning on detecting
the auxiliary atom in state $\ket{g}$ at the end of the gate thus rules out the
possibility of any dissipative quantum jumps having occurred during the gate. As
a result, the conditional fidelity of the gate is greatly enhanced at the modest
cost of a finite but potentially low failure probability.

We now analyze the performance of the gates and
derive the success probabilities, gate times and gate errors (see Tab.~S1 in \cite{SM})  
The Hamiltonian in a proper rotating frame is (see \figref{fig:levels})
\begin{eqnarray}
\hat{H}&=&\Delta_{E}\ket{E}\bra{E}+g_{f}(\hat{a}\ket{E}\bra{f}+H.c)+\hat{V}+\hat{V}^{\dagger} \nonumber \\
&&+\sum_{k}\Delta_{e}\ket{e}_{k}\bra{e}+g(\hat{a}\ket{e}_{k}\bra{1}+H.c), 
\end{eqnarray}          
where $k$ labels the qubit atoms ($\hbar=1$), $2\hat{V}=\Omega\ket{E}\bra{g}$ and we have
assumed that all couplings ($g,\Omega$) are real. We have defined
$\Delta_{E}=\omega_{E}-\omega_{g}-\omega_{L}$, and
$\Delta_{e}=\omega_{e}-\omega_{g}-\omega_{L}+\omega_{f}-\omega_{1}$, where
$\omega_{L}$ is the laser frequency and otherwise $\omega_{x}$ is the frequency
associated with level $x$. We describe the cavity decay and atomic spontaneous
emission with Lindblad operators so that $\hat{L}_{0}=\sqrt{\kappa}\hat{a}$
corresponds to the cavity decay,  $\hat{L}_{f}=\sqrt{\gamma_{f}}\ket{f}\bra{E}$
to the decay of the excited state of the auxiliary atom and
$\hat{L}_{k}=\sqrt{\gamma}\ket{\tilde{o}}_{k}\bra{e}$ describes the decay of the
excited qubit states to some arbitrary ground state $\ket{\tilde{o}}$. The
nature of $\ket{\tilde{o}}$ is not important for the dynamics of the gates and
it may or may not coincide with $\ket{0}$ or $\ket{1}$.

We assume a weak driving pulse justifying for a perturbative treatment of $\hat{V}$ using the formalism of Ref.~\cite{Florentin}. In the perturbative description
we adiabatically eliminate the coupled excited states of the atoms and the
cavity (assuming $\Omega^{2}/\Delta_{E}\!\ll\!\Delta_{E}$ and $\Omega\!\ll\!g$), which leads to an energy shift of the ground states but otherwise
conserves them since the Hamiltonian cannot connect different unexcited states
without decay. The dynamics are therefore described by an effective Hamiltonian,
$\hat{H}_{\text{eff}}=\ket{g}\bra{g}\sum_{n}\Delta_{n}\hat{P}_{n}$ where
\begin{equation}
\Delta_{n}=\mathrm{Re}\left\{\frac{-\frac{\Omega^{2}}{4\gamma}((\frac{\Delta_{e}}{\gamma}\!\!-\!\!i/2)i\!\!+\!\!2nC)}{(2\frac{\Delta_{e}}{\gamma}\!\!-\!\!i)((2\frac{\Delta_{E}}{\gamma}\!\!-\!\!i)i/4\!\!+\!\!C)\!\!+\!\!(2\frac{\Delta_{E}}{\gamma}\!\!-\!\!i)nC}\right\}\label{eq:detunings}
\end{equation} 
and $\hat{P}_{n}$ projects on the states with $n$ qubits in state $\ket{1}$. For
simplicity, we have assumed that the auxiliary atom is identical to the qubit
atoms such that $g_{f}=g$ and $\gamma_{f}=\gamma$ (see \cite{SM} for a more
general treatment) and we have defined the cooperativity $C=g^{2}/\gamma\kappa$.
We consider the limit $C\gg1$ and from Eq.~\eqref{eq:detunings} we find that the
energy shift, in the case when all qubit atoms are in $\ket{0}$, becomes very
small $\Delta_{0}\sim\Delta_{E}\Omega^{2}/(16\gamma^{2} C^{2})\rightarrow 0$, i.e.,
we drive into a zero energy dark state as mentioned in the description above.
On the contrary, for $n>0$, the $C$ in the nominator of $\Delta_{n}$ reflects that
the coupling of the qubit atoms shifts the cavity resonance and
as a result an AC stark shift of $\sim\Omega^{2}/\Delta_{E}$ is introduced.
Furthermore, we find that in the effective evolution, errors caused by
spontaneous emission or cavity decay ($\hat{L}_{0},\hat{L}_{f},\hat{L}_{k}$)
project the system out of the effective space into orthogonal subspaces, which
allows for an efficient error detection by measuring the ancilla atom. 

The dynamics described by $\hat{H}_{\text{eff}}$ can be used to implement a
Toffoli gate. Assuming the qubit atoms to be on resonance $(\Delta_{e}=0)$ and having
$\Delta_{E}\sim\gamma\sqrt{C}$ gives energy shifts
$\Delta_{n>0}\sim\Omega^{2}/(4\gamma\sqrt{C})$ while
$\Delta_{0}\sim\mathcal{O}(\Omega^{2}/C^{3/2})$. Hence, $\ket{00...0}$ is the only state, which
remains unshifted and we can choose a gate time of
$t_{\text{T}}\sim4\pi\sqrt{C}\gamma/\Omega^{2}$ to make a Toffoli gate. By
conditioning on measuring the auxiliary atom in state $\ket{g}$ at the end of
the gate, the detectable errors from cavity decay and spontaneous emission only
reduce the success probability instead of reducing the fidelity. Consequently, the fidelity becomes
limited by more subtle, undetectable errors (see Ref.~\cite{SM}). The dominant
error originates from the qubit dependent decay rate, $\Gamma_{n}$, of $\ket{g}\to\ket{f}$. As we demonstrate in
Ref.~\cite{SM}, this leads to a fidelity lower bounded by $1-F \lesssim 0.3/C$,
with a success probability of $P_{\text{s}} \sim 1 - 3/\sqrt{C}$. Thus is a substantial improvement over the leading error in the
case of deterministic cavity-assisted gates. For generic states, the fidelity can even be markedly higher, and improving
with increasing particle number $N$ \cite{SM} 

\begin{figure} 
\centering
\subfloat {\label{fig:prob_time}\includegraphics[width=0.25\textwidth,height=1.5in]{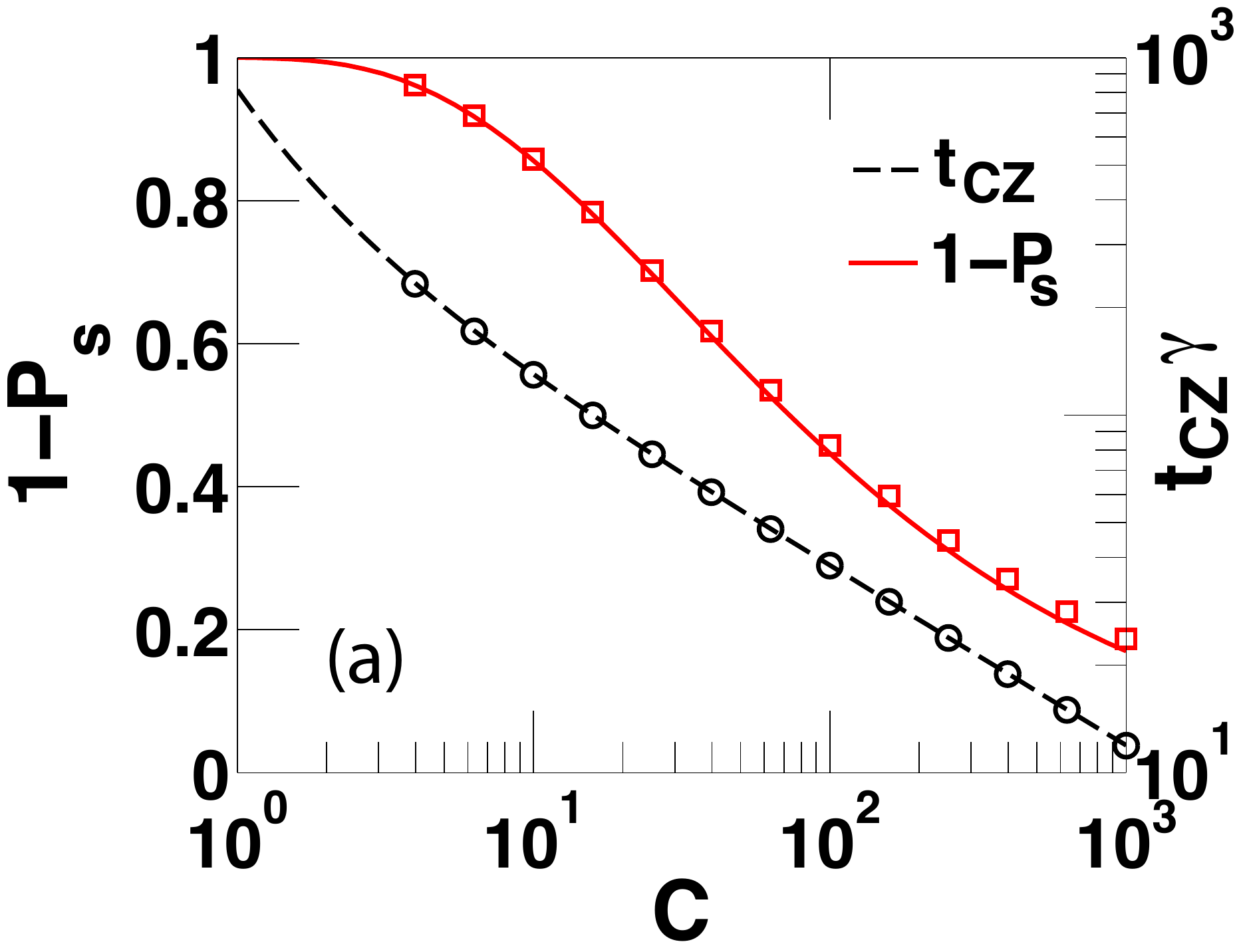}} 
\subfloat{\label{fig:fidel_time}\includegraphics[width=0.24\textwidth,height=1.49in]{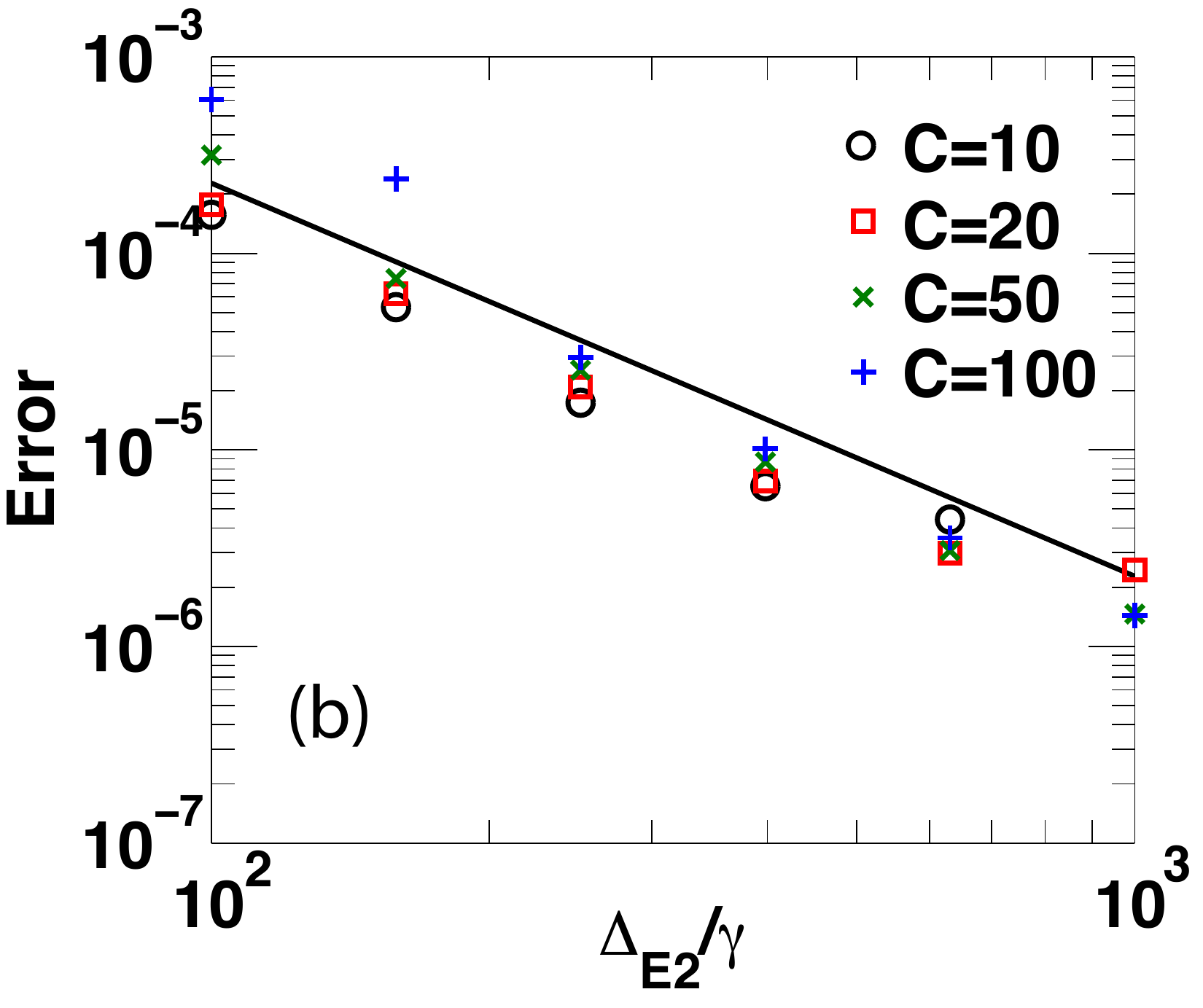}}
\caption{(Color online)(a) Failure probability ($1-P_{s}$ - left axis) and gate time ($t_{\text{CZ}}$ -right axis) as a function of the cooperativity ($C$) for the CZ gate. The gate time is in units of the inverse linewidth $1/\gamma$ of the qubit atoms. We have assumed a driving of $\Omega=\sqrt{C}\gamma/4$. (b) Gate error as a function of the detuning $\Delta_{E2}$ in the two-photon-driven CZ-gate for $C=10,20,50$, and $100$. We have assumed that $\Omega_{\text{\text{MW}}}=4\gamma C^{1/4}$ and that $\gamma_{g}=\gamma$. The gate error decreases as $\gamma^{2}/\Delta_{E2}^{2}$ and is independent of $C$.  We have assumed $\Omega\sim\Delta_{E2}/8$ resulting in a gate time $\sim400/\gamma$. Solid/dashed lines are analytical results and symbols are numerical simulations (see \cite{SM}). For both plots, we have assumed $\kappa=100\gamma$.}
\label{fig:prob}
\end{figure}

In the special case of only two qubits, the
Toffoli gate is referred to as a CZ-gate, and in this case, we can even improve the gate to have an arbitrarily small error
by combining it with single qubit rotations. For the general Toffoli gate
discussed above, we needed $\Delta_e=0$ to ensure the correct phase
evolution, but making the single qubit transformations $\ket{0}\to e^{-i\Delta_{0}t/2}\ket{0}$
and $\ket{1}\to e^{-i(\Delta_{1}-\Delta_{0})t/2}\ket{0}$, at the end of a
driving pulse of length 
$t_{\text{CZ}}=\abs{\pi/(\Delta_{2}-2\Delta_{1}+\Delta_{0})}$, ensures the right
phase evolution of the CZ-gate without any constraints on $\Delta_{e}$. Hence, it is possible to tune $\Delta_{e}$ to eliminate the
detrimental effect of having a qubit dependent decay rate. Choosing
$\Delta_{E}=\frac{\gamma}{2}\sqrt{4C+1}$ and
$\Delta_{e}=\frac{1}{2}C\gamma^{2}/\Delta_{E}$ ensures $\Gamma_0=\Gamma_1=\Gamma_2$, and thus removes
all dissipative errors from the heralded gate. The conditional error is then
limited only by non-adiabatic effects, that can in principle be made arbitrarily
small by reducing the driving strength. The success probability is
$1-P_{\text{s}}\sim6/\sqrt{C}$ in the limit $C\gg1$ (see
\figref{fig:prob_time}).
We thus have a heralded two qubit gate with arbitrarily small error with a
success probability that can approach 1 (it is possible to decrease the scaling factor of the probability from $\sim6$ to $\sim3.4$ at the expense of an error scaling as $1/C$ by tuning $\Delta_{E},\Delta_{e}$). 

We now consider the gate time. The gate time of the Toffoli gate is
$t_{\text{T}}\sim4\pi\sqrt{C}\gamma/\Omega^{2}$ and for the CZ-gate we have
$t_{\text{CZ}}\sim15\pi\sqrt{C}\gamma/(2\Omega^{2})$ for $C\gg1$. Since
$t_{\text{CZ}}>t_{\text{T}}$ we focus on $t_{\text{CZ}}$. The gate time is set
by the strength ($\Omega$) of the driving pulse, which is limited by non-adiabatic errors. This is investigated in the supplemental material where we also verify our analytical results numerically~\cite{SM}. Assuming realisitc parameters of $\kappa=100\gamma$ \cite{thompson,Tiecke}, we find that a driving of $\Omega=\sqrt{C}\gamma/4$ 
keeps the non-adiabatic error of the gate below $4\cdot10^{-5}$ for $C\leq1000$. The gate times decreases as $1/\sqrt{C}$ as shown in \figref{fig:prob_time}. For a cooperativity of 
$100$ the gate time is $\approx1$ $\mu$s  for typical atomic decay rates.  

So far, we have assumed a model where there is no decay from
$\ket{E}\to\ket{g}$. In real atoms, there will, however, always be some decay
$\ket{E}\to\ket{g}$ with a decay rate $\gamma_{g}>0$. The result of such an
undetectable decay is that both the CZ-gate and the Toffoli gate will have an
error $\sim \gamma_{g}/(\gamma\sqrt{C})$. To make this error small, it is thus essential to suppress the
branching ratio $\gamma_{g}/\gamma$. Below we show how to suppress $\gamma_{g}$ by driving the $\ket{g}\to\ket{E}$
transition with a two photon process. As a result, we realize a CZ gate with an
error arbitrary close to zero and a Toffoli gate with an error scaling  as $1/C$
even for a realistic atomic system.

\begin{figure} 
\centering
\includegraphics[width=0.5\textwidth, height=1.6in]{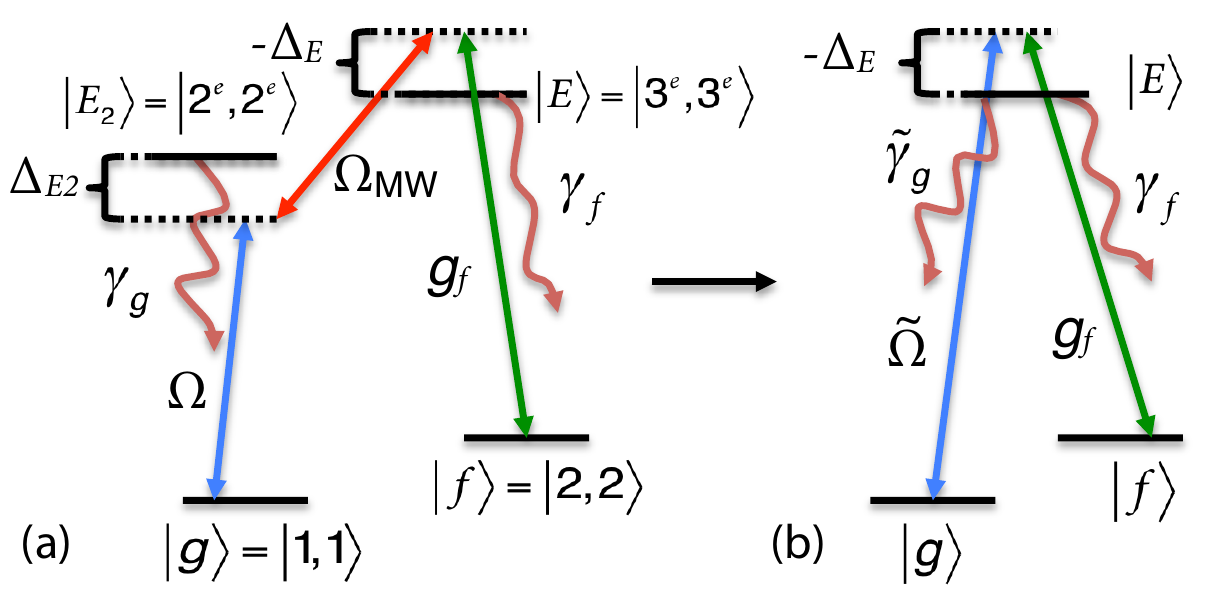}
\caption{(Color online) (a) Level
structure of the auxiliary atom and the transitions driven by a weak laser ($\Omega$), a microwave field ($\Omega_{\text{MW}}$) and the cavity ($g_{f}$). We assume that
$\ket{E}\leftrightarrow\ket{f}$ is a closed transition and, for simplicity, we
also assume that $\ket{E_{2}}\leftrightarrow\ket{g}$ is a closed transition but
this is not a necessity. Here $\ket{r^{(e)},r^{(e)}}$ with $r=1,2,3$ refers to
how the atom may be realized in the $(5^{2}P_{3/2})$ states
$\ket{F^{(e)}=r,m^{(e)}=r}$ $5^{2}S_{1/2}$ of Rb${}^{87}$. (b) Effective
three-level atom realized by mapping the two-photon drive to give an effective
decay rate $\tilde{\gamma}_{g}$ and an effective drive $\tilde{\Omega}$. }
\label{fig:control}
\end{figure}

Specifically we think of a level structure for the auxiliary atom, shown in
\figref{fig:control},
where we still assume $\ket{E}\leftrightarrow\ket{f}$ to be a closed transition.
For simplicity, we have also assumed $\ket{E_{2}}\leftrightarrow\ket{g}$ to be a
closed transition. Such a level structure could,
e.g. be realized in  ${}^{87}$Rb  as shown in \figref{fig:control}.  We assume
that a microwave field couples the two excited states such that we can have a
two photon transition from $\ket{g}\to\ket{E}$ and that $\Omega$ is small,
allowing for a perturbative treatment of the coupling. Thus we can map
the system to a simple three-level atom with levels $\ket{g},\ket{E}$ and
$\ket{f}$ and a decay rate $\tilde{\gamma}_{g}$ and drive $\tilde{\Omega}$
between $\ket{g}$ and $\ket{E}$, determined by the two photon driving process as
shown in \figref{fig:control}. The dynamics are thus similar to what we have
already described for the simple three level atom except that we have the extra
decay $\tilde{\gamma}_{g}$ that introduces an error in the gates
$\sim(\tilde{\gamma}_{g}/\gamma)/\sqrt{C}$, as previously described. In the
limit $C\gg1$, we find
$\tilde{\gamma}_{g}/\gamma\sim\frac{\gamma_{g}\Omega_{\text{MW}}^{2}}{4\gamma\Delta_{E2}^{2}}$.
Thus by increasing $\Delta_{E2}$, we can in principle make these errors
arbitrarily small. The error of the CZ-gate for different $\Delta_{E2}$ is shown
in \figref{fig:fidel_time}, assuming an initial state of
$(\ket{0}+\ket{1})^{\otimes 2}$. Note that in order to prevent an increasing
scattering probability of level $\ket{E2}$, we need to have $\Omega_{MW}\propto
C^{1/4}$ resulting in a gate error that is independent of the cooperativity.
\cite{SM}.
The success probability and time of the gates are the same as before with
$\Omega\to\tilde{\Omega}\sim\frac{\Omega_{\text{MW}}\Omega}{2\Delta_{E2}}$. With
similar considerations about the validity of our perturbation as
before, we find that for realistic parameters, we can use $\Omega=\Delta_{E2}/8,
\Omega_{\text{MW}}\sim4\gamma C^{1/4}$ resulting in a gate time of $\sim10$ $\mu$s for typical atomic decay rates and $C\lesssim1000$~\cite{SM}.   

As an example implementation, we consider ultra-cold ${}^{87}$Rb atoms coupled to nanophotonic cavities~\cite{thompson,Tiecke}. There are some additional errors originating from the extra states in the ${}^{87}$Rb atoms in this case. In Ref. \cite{SM}, we treat these errors and find that with a detuning of $\Delta_{E2}=100\gamma$ and a cooperativity of $C\approx100$, a heralded CZ gate with $\sim67\%$ success probability and a heralded error of $\approx 10^{-3}$ can be realized
in $\approx10$ $\mu$s time. This justifies neglecting atomic decoherence which is typically much slower.  
Alternatively the gate can be implemented with atom-like solid-state qubits such as NV and SiV centers in diamond~\cite{phystoday}. These systems can exhibit closed transitions and long-lived electronic spin states which are the essential requirement for the gate~\cite{togan}, while high cooperativities are possible in  diamond nanocavities~\cite{burek}. A particular advantage of such system is the long-lived nuclear spin degrees of freedom, which allows each of the color centers  to act as a multi-qubit quantum network node~\cite{maurer}.   By entangling electronic spins via the heralded gate, a high-fidelity, fully deterministic gate can subsequently be performed on qubits stored in nuclear spins~\cite{Anders2prl}.

As a particular application, we consider a quantum repeater where entanglement is first created in small segments (links), which are subsequently connected using entanglement swapping~\cite{briegel}. By organizing the repeater in a tree structure, the probabilistic nature of the gate can be efficiently circumvented. The success rate of distributing entanglement across the total distance L, scales as $\sim (L/L_{0})^{1-\log_{2}(3/p)}$, where $p<1$ is the success probability of the swap, $L$ is the total distribution distance and $L_{0}$ is the length between the links~\cite{repeater} (note that in the limit $p\to1$, the above expression underestimates the rate, e.g., for $p=1$ the actual rate is $\sim3$ times faster for 128 links). This is a substantial improvement over direct transmission where the success rate scales exponentially with $L$. For a realization with nuclear spin memories where the swap can be performed deterministically the rate can scale even better as $\sim\log_{2}(L/L_{0})^{-1}$. In order to maintain the favorable scaling without resorting to time consuming purification, the total number of links, $N_{\text{max}}$ should be kept below $N_{\text{max}}\sim-\ln(F_{\text{final}})/(\epsilon_{0}+\epsilon_{g})$, where $F_{\text{final}}$ is the required fidelity of the final distributed pair and $\epsilon_{0},\epsilon_{g}\ll1$ are the errors of the initial entanglement generation and the entanglement swapping respectively. Thus, it is essential that the errors are kept small, which can be obtained with the heralded gate.   

In conclusion, we have introduced a heralded two-qubit quantum
gate with a conditional fidelity arbitrarily close to unity and an $N$-qubit Toffoli gate
with an error scaling as $1/C$. The gates have a
built-in error detection process, which removes the necessity of extracting the
error by the more complicated process of entanglement purification or quantum
error correction. Our gate is designed for the specific case of optical
cavities, and allows exploiting realistic systems for quantum communication, even though the error rate would inhibit this with deterministic gates. Similar
advantages can be realised in other systems such as those based on circuit QED, where certain errors could be
heralded and thus alleviate the daunting requirements of fault tolerant
computation. 

We thank Jeff Thompson for helpful discussions and gratefully acknowledge the support from the Lundbeck Foundation, NSF, CUA, DARPA, AFOSR MURI, and ARL. The research leading to these results has received funding from the European Research Council
under the European Union's Seventh Framework Programme (FP/2007-2013) / ERC
Grant Agreement n. 306576 and through SIQS (grant no. 600645) and QIOS (grant
no. 306576).

\newpage

\renewcommand{\theequation}{S\arabic{equation}}
\renewcommand{\thefigure}{S\arabic{figure}}
\renewcommand{\thetable}{S\arabic{table}}

\newcommand{\bel}{\begin{equation}}
\newcommand{\eel}{\end{equation}}
\newcommand{\be}{\begin{equation*}}
\newcommand{\ee}{\end{equation*}}
\newcommand{\bal}{\begin{eqnarray}}
\newcommand{\eal}{\end{eqnarray}}
\newcommand{\ba}{\begin{eqnarray*}}
\newcommand{\ea}{\end{eqnarray*}}

\newcommand{\no}{\noindent}

\newcommand{\refeq}[1]{Eq.~(\ref{#1})}
\newcommand{\reffig}[1]{Fig.~\ref{#1}}

\newcommand{\ev}[1]{\langle #1 \rangle}
\newcommand{\Ev}[1]{\left\langle #1 \right\rangle}
\newcommand{\Ket}[1]{\left| #1 \right\rangle}
\newcommand{\Bra}[1]{\left\langle #1 \right|}
\newcommand{\+}{^\dagger}
\newcommand{\s}{^\ast}
\renewcommand{\ss}{^{\ast\ast}}
\newcommand{\Tr}{\text{Tr}\,}

\newcommand{\DD}{\mathcal{D}}
\newcommand{\PP}{\mathcal{P}}
\newcommand{\pk}[1]{\textcolor[rgb]{1,0,0}{#1}}

\onecolumngrid

\section{Supplemental material}

\onecolumngrid

This supplemental material to the article "Heralded quantum gates with integrated error detection in optical cavities" describes the details of the perturbation theory and the derivation of the effective Hamiltonian $\hat{H}_{\text{eff}}$ and effective Lindblad operators. We describe the situation both with and without a two-photon drive. Furthermore, we present the results of a numerical simulation of the full dynamics of the gates to verify the results found with perturbation theory and address the question of how strong a drive we can allow for. In the end, this determines the gate time as described in the article. Finally we discuss of the additional errors described in the final part of the article. 

\section{Perturbation theory}

We will now give the details of the perturbation theory and the derivation of the effective operators together with the success probabilitites, gate times and gate errors (see \tabref{tab:table1}). Our perturbation theory is based on the effective operator formalism described in Ref.~\cite{Florentin}.

\begin{table} [h]
\centering
\begin{tabular}{|c|c|c|c|c|}
\hline
Gate & Origin of error & Error & Probability & Time  \\ \hline CZ-gate & \specialcell{$\gamma_{g}=0$
\\$\gamma_{g}>0$} & \specialcell{$0$\\$\sim
\frac{\gamma_{g}}{\gamma\sqrt{C}}$} & $\sim 1-\frac{6}{\sqrt{C}}$ & $\sim\frac{15\pi\sqrt{C}\gamma}{2\Omega^{2}}$\\ \hline Toffoli &
\specialcell{$\Gamma_{i}\neq\Gamma_{j}$\\$\gamma_{g}>0$} &
\specialcell{$\lesssim\!\frac{0.3}{C}\quad$\\$\sim\!\!
\frac{\gamma_{g}}{\gamma\sqrt{C}}$} & $\sim 1-\frac{3}{\sqrt{C}}$ & $\sim\frac{4\pi\sqrt{C}\gamma}{\Omega^{2}}$ \\ \hline
\end{tabular}
\caption{The errors, success probabilities and gate times of the $N$-qubit Toffoli gate and the $CZ$-gate considered
in the article. Note that the branching fraction $\gamma_{g}/\gamma$ can be made arbitrarily small using a far detuned two-photon driving as explained in below. $\Gamma_{i}$ is the rate of detectable errors for the qubit state with $i$ qubits in state $\ket{1}$. The success probability of the CZ-gate can be increased at the expense of an error scaling of $1/C$ as explained in the article.}
\label{tab:table1}
\end{table}  

\begin{figure} [H]
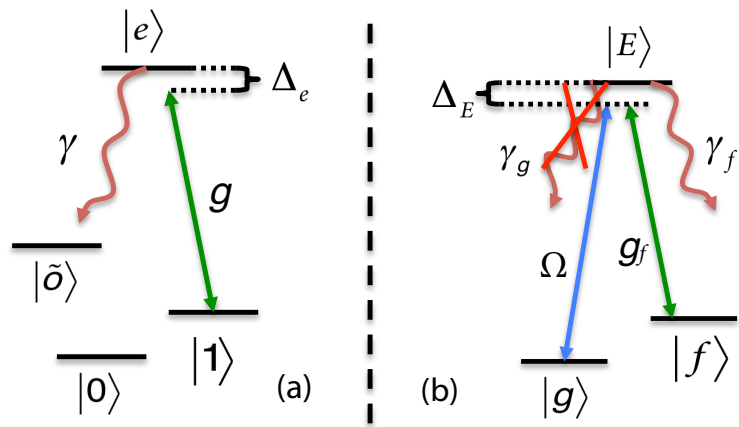

\centering
\subfloat {\label{fig:figureS1a}\includegraphics[width=0.31\textwidth]{figureS1a}}
\subfloat {\label{fig:figureS1b}\includegraphics[width=0.3\textwidth]{figureS1b}}
\caption{(a) Level structure of the qubit atoms. Only state $\ket{1}$ couples to the cavity and we assume that the excited level decays to some level $\ket{\tilde{o}}$, possible identical to $\ket{f}$ or $\ket{0}$. (b) Level structure of the auxiliary atom and the transitions driven by the weak laser ($\Omega$) and the cavity ($g_{f}$). We allow for some decay from $\ket{E}\to\ket{g}$ with decay rate $\gamma_{g}$. }
\label{fig:figureS1}
\end{figure}

First, we treat the simplest situation where the auxiliary atom is directly driven to an excited state $\ket{E}$ by a weak classical drive $\Omega$ as shown in \figref{fig:figureS1} (reproduced from Fig. 1 in the article). Note that we allow for some decay from $\ket{E}\to\ket{g}$ with decay rate $\gamma_{g}$ as opposed to the situation in the article. We will later consider the situation where this decay rate is suppressed using a two-photon drive. The level structure of the qubit atoms are also shown in \figref{fig:figureS1}. 

The Hamilton describing the system in a proper rotating frame is given by Eqs. (1)-(3) in the article and is reproduced here 
\begin{eqnarray} \label{eq:hamil1}
\hat{H}&=&\hat{H}_{e}+\hat{V}+\hat{V}^{\dagger}, \\
\hat{H}_{e}&=&\Delta_{E}\ket{E}\bra{E}+g_{f}(\hat{a}\ket{E}\bra{f}+H.c) \nonumber \\
&&+\sum_{k}\Delta_{e}\ket{e}_{k}\bra{e}+g(\hat{a}\ket{e}_{k}\bra{1}+H.c), \\
\hat{V}&=&\frac{\Omega}{2}\ket{E}\bra{g},
\end{eqnarray}          
where we have assumed for simplicty that all couplings ($g,\Omega$) are real and $k$ labels the qubit atoms ($\hbar=1$). We have defined $\Delta_{E}=\omega_{E}-\omega_{g}-\omega_{L}$, and $\Delta_{e}=\omega_{e}-\omega_{g}-\omega_{L}+\omega_{f}-\omega_{1}$ where $\omega_{L}$ is the laser frequency and otherwise $\omega_{x}$ is the frequency associated with level $x$. Note that we assume the cavity frequency to be $\omega_{c}=\omega_{L}+\omega_{g}-\omega_{f}$ such that we are on resonance with the $\ket{g}\to\ket{E}\to\ket{f}$ two-photon transition. 

The dissipation in the system is assumed to be described by Lindblad operators such that $\hat{L}_{0}=\sqrt{\kappa}\hat{a}$ describes the cavity decay with decay rate $\kappa$, $\hat{L}_{g}=\sqrt{\gamma_{g}}\ket{g}\bra{E}$, and $\hat{L}_{f}=\sqrt{\gamma_{f}}\ket{f}\bra{E}$ describes the decay of the auxiliary atom and $\hat{L}_{k}=\sqrt{\gamma}\ket{\tilde{o}}_{i}\bra{e}$ describes the decay of the qubit atoms ($k=1,2\ldots N$). As described in the article $\ket{\tilde{o}}$ may or may not coincide with $\ket{0}$ or $\ket{1}$. Assuming that $\Omega$ is weak ($\Omega^{2}/\Delta_{E}\ll\Delta_{E}$ and $\Omega\ll g$), we can treat the driving as a perturbation to the system. As shown in Ref.~\cite{Florentin}, the dynamics of the system is then governed by an effective master equation of the form
\begin{equation}
\dot{\rho}=i\left[\rho,\hat{H}_{\text{eff}}\right] + \sum_{x}\hat{L}_{x}^{\text{eff}}\rho(\hat{L}_{x}^{\text{eff}})^{\dagger}-\frac{1}{2}\left((\hat{L}_{x}^{\text{eff}})^{\dagger}\hat{L}_{x}^{\text{eff}}\rho+\rho(\hat{L}_{x}^{\text{eff}})^{\dagger}\hat{L}_{x}^{\text{eff}}\right),
\end{equation}
where $\rho$ is the density matrix of the system, $\hat{H}_{\text{eff}}$ is an effective Hamiltonian, and  $L_{x}^{\text{eff}}$ are effective Lindblad operators with $x=0,g,f,k$.  The effective operators are found from:
\begin{eqnarray} \label{eq:supheff1}
\hat{H}_{\text{eff}}&=&-\frac{1}{2}\hat{V}^{\dagger}\left(\hat{H}_{\text{NH}}^{-1}+(\hat{H}_{\text{NH}}^{-1})^{\dagger}\right)\hat{V} \\
\hat{L}_{x}^{\text{eff}}&=&\hat{L}_{x}\hat{H}_{\text{NH}}^{-1}\hat{V},  \label{eq:supleff1}
\end{eqnarray}
where 
\begin{equation}
\hat{H}_{\text{NH}}=\hat{H}_{e}-\frac{i}{2}\sum_{x}\hat{L}_{x}^{\dagger}\hat{L}_{x}, 
\end{equation}
is the no-jump Hamiltonian. The Hilbert space of the effective operators can be described in the basis of $\left\{\ket{g},\ket{f}\right\}$ of the auxiliary atom and the states $\left\{\ket{0},\ket{1},\ket{\tilde{o}}\right\}$ of the qubit atoms. To ease the notation, we define the projection operators $\hat{P}_{n}$ which projects on to the states with $n$ qubits in state $\ket{1}$. From Eq.~\eqref{eq:supheff1} and \eqref{eq:supleff1} we then find:
\begin{eqnarray}
\hat{H}_{\text{eff}}&=&\sum_{n=0}^{N}\frac{-\Omega^{2}}{4\gamma}\mathrm{Re}\left\{\frac{i\tilde{\Delta}_{e}/2+nC}{\tilde{\Delta}_{e}(i\tilde{\Delta}_{E}/2+C_{f})+\tilde{\Delta}_{E}nC}\right\}\ket{g}\bra{g}\otimes \hat{P}_{n} \nonumber \\
&=&\sum_{n=0}^{N}\Delta_{n}\ket{g}\bra{g}\otimes \hat{P}_{n} \label{eq:supham2}\\
\hat{L}_{0}^{\text{eff}}&=&\sum_{n=0}^{N}\frac{1}{2\sqrt{\gamma}}\frac{\sqrt{C_{f}}\tilde{\Delta}_{e}\Omega}{\tilde{\Delta}_{e}(i\tilde{\Delta}_{E}/2+C_{f})+n\tilde{\Delta}_{E}C}\ket{f}\bra{g}\otimes \hat{P}_{n} \nonumber \\
&=&\sum_{n=0}^{N}r_{0,n}^{\text{eff}}\ket{f}\bra{g}\otimes \hat{P}_{n} \label{eq:subleff1} \\
\hat{L}_{g}^{\text{eff}}&=&\sum_{n=0}^{N}\frac{1}{2}\frac{(i\tilde{\Delta}_{e}/2+nC)\Omega}{\tilde{\Delta}_{e}(i\tilde{\Delta}_{E}/2+C_{f})+n\tilde{\Delta}_{E}C}\frac{\sqrt{\gamma_{g}}}{\gamma}\ket{g}\bra{g}\otimes \hat{P}_{n} \nonumber \\
&=&\sum_{n=0}^{N}r_{g,n}^{\text{eff}}\ket{g}\bra{g}\otimes \hat{P}_{n} \\
\hat{L}_{f}^{\text{eff}}&=&\sum_{n=0}^{N}\frac{1}{2}\frac{(i\tilde{\Delta}_{e}/2+nC)\Omega}{\tilde{\Delta}_{e}(i\tilde{\Delta}_{E}/2+C_{f})+n\tilde{\Delta}_{E}C}\frac{\sqrt{\gamma_{f}}}{\gamma}\ket{f}\bra{g}\otimes \hat{P}_{n} \nonumber \\
&=&\sum_{n=0}^{N}r_{f,n}^{\text{eff}}\ket{f}\bra{g}\otimes \hat{P}_{n} \\
\hat{L}_{k}^{\text{eff}}&=&\sum_{n=1}^{N}-\frac{1}{2\sqrt{\gamma}}\frac{\sqrt{C_{f}}\sqrt{C}\Omega}{\tilde{\Delta}_{e}(i\tilde{\Delta}_{E}/2+C_{f})+n\tilde{\Delta}_{E}C}\ket{f}\bra{g}\otimes\ket{\tilde{o}}_{k}\bra{1}\otimes\hat{P}_{n} \nonumber \\
&=&\sum_{n=1}^{N}r_{n}^{\text{eff}}\ket{f}\bra{g}\otimes\ket{\tilde{o}}_{k}\bra{1}\otimes\hat{P}_{n}, \label{eq:subleff2} 
\end{eqnarray}
where we have defined the cooperativities $C_{(f)}=g_{(f)}^{2}/\gamma\kappa$ for the qubit (auxiliary) atoms and the complex detunings $\tilde{\Delta}_{E}\gamma=\Delta_{E}-i\gamma_{f}/2$ and  $\tilde{\Delta}_{e}\gamma=\Delta_{e}-i\gamma/2$. Note that we have defined the parameters $r^{\text{eff}}_{0,n}, r^{\text{eff}}_{g,n},r^{\text{eff}}_{f,n}$ and $r_{n}^{\text{eff}}$ in Eqs. \eqref{eq:subleff1}-\eqref{eq:subleff2} to characterize the decays described by the Lindblad operators. Note that $r_{0}^{\text{eff}}=0$. In our calculations we parametrize the difference between the auxiliary atom and the qubit atoms by $C_{f}=\alpha C$ and $\gamma_{f}=\beta\gamma$ to easier treat the limit of $C\gg1$ that we are interested in. 

\subsection{Success probability and fidelity}
Eqs. \eqref{eq:subleff1}-\eqref{eq:subleff2} show that the effect of all Lindblad operators, except $\hat{L}_{g}^{\text{eff}}$, is that the state of the auxiliary atom is left in state $\ket{f}$. All these errors are thus detectable by measuring the state of the auxiliary atom at the end of the gate. For the heralded gates where we condition on measuring the auxiliary atom in state $\ket{g}$ at the end of the gates, these detectable decays therefore do not effect the fidelity of the gates but only the success probability.  The rate $\Gamma_{n}$ of the detectable decays for a state with $n$ qubits in state $\ket{1}$ is $\Gamma_{n}=\abs{r_{0,n}^{\text{eff}}}^{2}+\abs{r_{f,n}^{\text{eff}}}^{2}+\abs{r_{n}^{\text{eff}}}^{2}$ and 
assuming an initial qubit state described by density matrix $\rho_{qubit}$ the success probability of the gates is
\begin{equation}
P_{\text{success}}=\sum_{n=0}^{N}\text{Tr}\left\{e^{-\Gamma_{n}t_{\text{gate}}}\rho_{\text{qubit}}\hat{P}_{n}\right\},
\end{equation}
where $t_{\text{gate}}$ is the gate time and $\text{Tr}$ denotes the trace. 

Having removed the detrimental effect of the detectable errors by heralding on a measurement of the auxiliary atom the fidelity of the gates will be determined by more subtle, undetectable errors (see below). We define the fidelity, $F$ of the gate as
\begin{equation}
F=\frac{1}{P_{\text{success}}}\bra{\psi}\bra{g}\tilde{\rho}_{\text{qubit}}\ket{g}\ket{\psi},
\end{equation} 
where we have assumed that the ideal qubit state after the gate is a pure state $\ket{\psi}$ and $\tilde{\rho}_{qubit}$ is the actual density matrix of the qubits and the auxiliary atom after the gate operation. 

\subsection{$N$-qubit Toffoli gate}

As shown in the article, the effective Hamiltonian in Eq.~\eqref{eq:supham2} is sufficient to make a Toffoli gate by putting the qubit atoms on resonance ($\Delta_{e}=0$). We will now treat the worst case and average fidelities of the general Toffoli gate referred to in the article. The undetectable errors limiting the fidelities are the following.
\begin{itemize}
\item As described in the article the energy shifts of the coupled qubit states are all $\Delta_{n>0}\sim\Omega^{2}/(4\gamma\sqrt{C})$ in the limit $C\gg1$. However, to higher order in $C$, we find corrections on the order $\mathcal{O}(\Omega^{2}/C^{3/2})$ to the energy shifts, which depend on the number of qubits that couples. The gate time of the Toffoli gate is $t_{\text{T}}\sim4\pi\sqrt{C}\gamma/\Omega^{2}$ and consequently, the higher order corrections give uneven phase shifts on the order of $\mathcal{O}(C^{-1})$ for the coupled qubit states at the end of the gate. This leads to a phase error in the fidelity of $\mathcal{O}(C^{-2})$. 
\item The difference between the rates of detectable errors ($\Gamma_{n}$) for different qubit states changes the relative weight of the qubit states during the gate. This error wil be $\mathcal{O}(C^{-1})$ as shown below. 
\item For $\gamma_{g}>0$ the undetectable decay from $\ket{E}\to\ket{g}$ in the auxiliary atom will destroy the coherence between the qubit states. We find that this error will be $\sim \frac{\gamma_{g}}{\gamma\sqrt{C}}$. For now, we will assume that $\gamma_{g}=0$ and thus ignore this error since we will show that we can suppress the branching fraction $\gamma_{g}/\gamma$ arbitrary close to zero by having a two photon driving.  
\end{itemize}  

\begin{figure} 
\centering
\subfloat {\label{fig:toffolia}\includegraphics[width=0.4\textwidth]{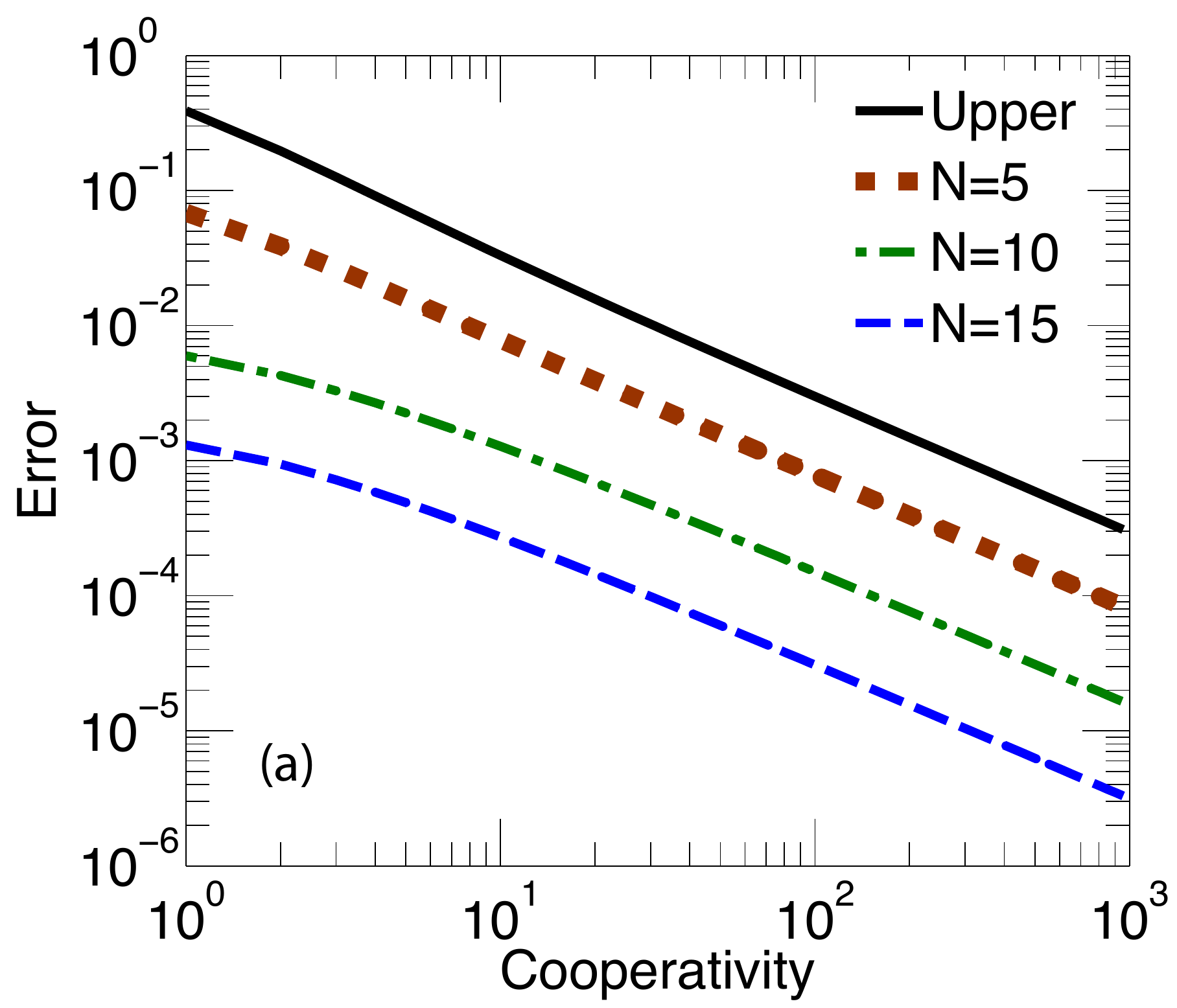}} 
\subfloat{\label{fig:toffolib}\includegraphics[width=0.39\textwidth]{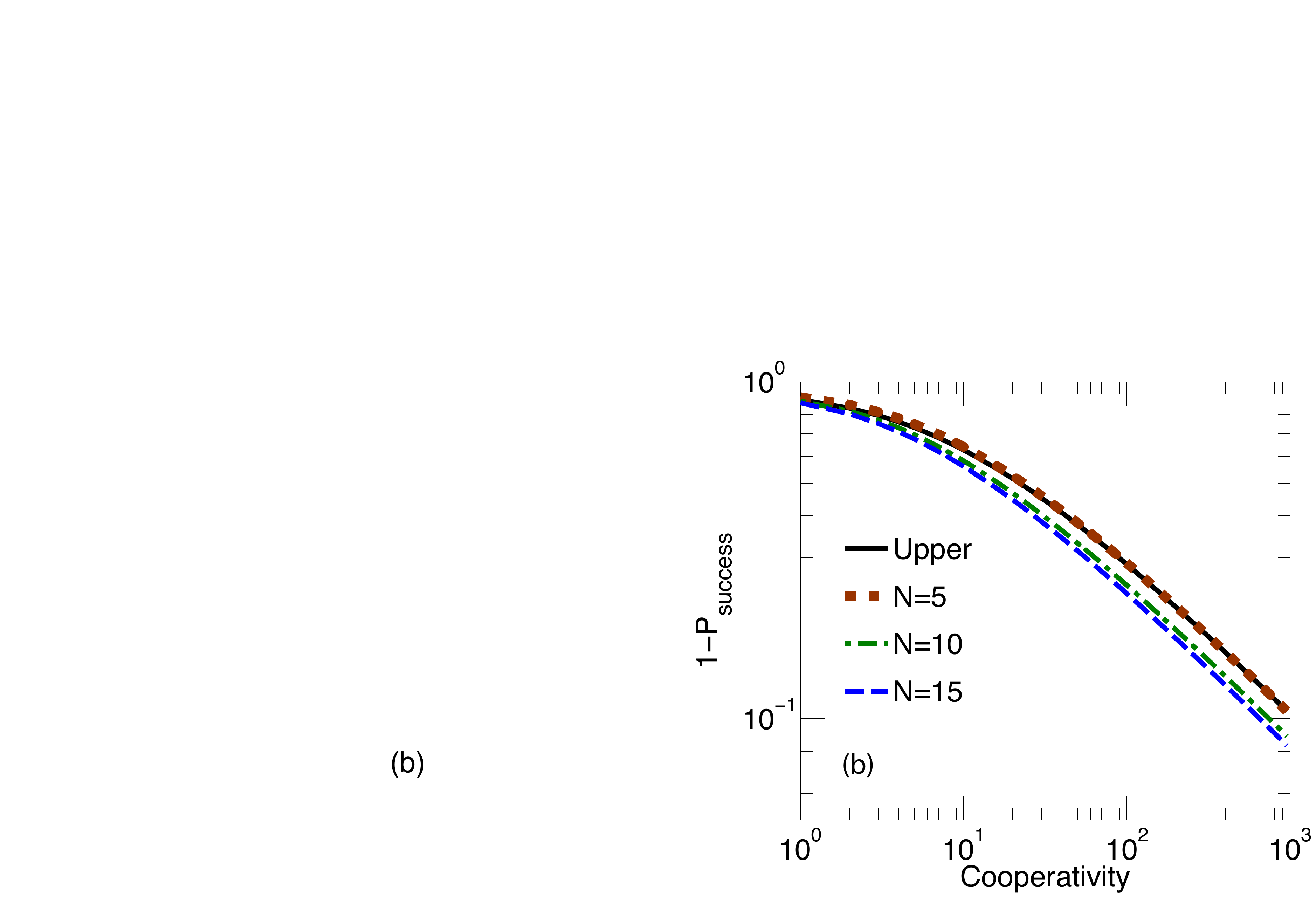}}
\caption{(Color online) (a) Gate error of the Toffoli for different initial states as a function of cooperativity. We have plotted the generic error for $N=5,10$, and 15 and the upper bound of the error. Note that the generic error decreases as $N$ increases. We have fixed $\Delta$ such that $\Gamma_{0}=\Gamma_{1}$ and have assumed that $\alpha=\beta=1$. (b) The failure probabilities $1-P_{\text{success,up}}$ and  $1-P_{\text{success,gen}}$ as a function of cooperativity. $1-P_{\text{success,gen}}$ is plotted for $N=5,10,15$. We have used the same assumptions as in (a). In general, the failure probability only have a weak dependence on $N$. Note that the line for $1-P_{\text{success, gen}},N=5$ coincides with  $1-P_{\text{success,up}}$.}
\label{fig:toffoli}
\end{figure}

Assuming that $\gamma_{g}=0$, the dominating source of error limiting the performance of the Toffoli gate is thus the difference between the rates of the detectable errors for the qubit states. We tune $\Delta_{E}$ such that $\Gamma_{0}=\Gamma_{1}$ and the largest difference between the detectable errors is thus between the completely uncoupled state and the state with all qubit atoms in state $\ket{1}$. As a result, we can find an upper bound on the fidelity of the $N$ qubit Toffoli gate, considering an initial state $\ket{0}^{\otimes N}+\ket{1}^{\otimes N}$ in the limit $N\to\infty$ because this state experiences the largest difference between the number of coupled and uncoupled qubits. We find that the upper bound on the fidelity and the corresponding success probability is
\begin{eqnarray}
F_{\text{up}}&\sim&1-\frac{\pi^{2}\alpha}{16(\alpha+\beta)}\frac{1}{C}\\
P_{\text{success,up}}&\sim&1-\frac{(\alpha+2\beta)\pi}{2\sqrt{\alpha}\sqrt{\alpha+\beta}} \frac{1}{\sqrt{C}}.
\end{eqnarray}.   
In general, the fidelity of the gate will, however, be larger than what is suggested above. Considering a generic input state  $(\ket{0}+\ket{1})^{\otimes N}$ with the same parameters as above, we find
\begin{eqnarray}
F_{\text{gen}}&\sim&1-k(N)\frac{\alpha \pi^{2}}{\alpha+\beta}\frac{1}{C} \\
P_{\text{success,gen}}&\sim&1-\frac{(d(N)\alpha+2\beta)\pi}{2\sqrt{\alpha}\sqrt{\alpha+\beta}} \frac{1}{\sqrt{C}},
\end{eqnarray}
where $k(N),d(N)$ are scaling factors which depend on the number of qubits $N$. We calculate $k(N)$ and $d(N)$ numerically for $N=1-100$ using the perturbation theory and find that that they both decrease with $N$ (see \figref{fig:toffoli}). The upper bounded and generic fidelities and corresponding success probabilities are shown in \figref{fig:toffoli} for different number of qubits, $N$.  
As $N$ increases we obtain higher generic fidelity, whereas the success probability is almost independent of $N$. 

\subsection{CZ-gate}

In the special case of only two qubits the Toffoli gate is referred to as a control-phase (CZ) gate. As shown in the article, we can, in this case, completely remove the errors from the gate by choosing the detunings $\Delta_{E}$ and $\Delta_{e}$ such that $\Gamma_{0}=\Gamma_{1}=\Gamma_{2}$ and combining it with single qubit rotations we can ensure the right phase evolution. In the general case where $\alpha,\beta\neq1$, the detunings $\Delta_{e}$ and $\Delta_{E}$ are 
\begin{eqnarray}
\Delta_{E}&=&\frac{\gamma}{2}\sqrt{\beta}\sqrt{4\alpha C+\beta} \\
\Delta_{e}&=&\frac{\alpha C\gamma^{2}}{2\Delta_{E}}. 
\end{eqnarray}
The success probability of the gate is then
\begin{equation}
P_{\text{success}}\simeq1-\pi\frac{8\beta^{2}+6\beta\alpha+\alpha^{2}}{8\beta^{3/2}\sqrt{\alpha}}\frac{1}{\sqrt{C}},
\end{equation} 
and we find that the gate time is $t_{\text{CZ}}\simeq\frac{\gamma\pi\sqrt{\alpha}(\alpha+2\beta)(\alpha+4\beta)}{2\beta^{3/2}\Omega^{2}}\sqrt{C}$ in the limit $C\gg1$.
\subsection{Two-photon driving}

\begin{figure} [H]
\centering
\includegraphics[width=0.7\textwidth, height=2.5in]{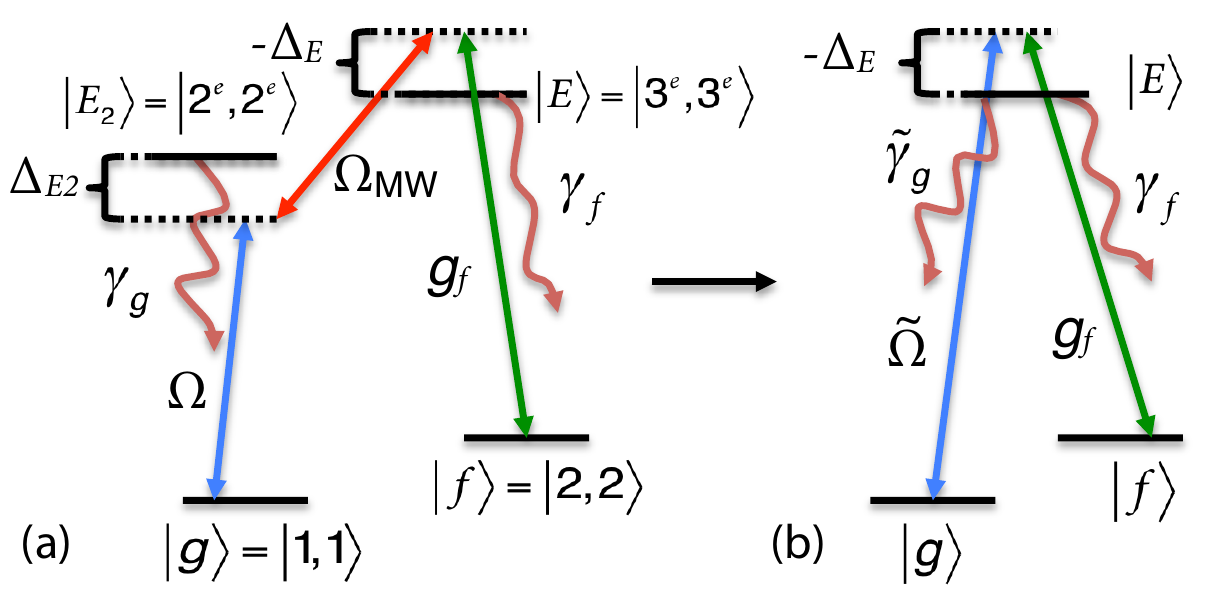}
\caption{(a) Level structure of the auxiliary atom and the transitions driven by a weak laser ($\Omega$), a microwave field ($\Omega_{\text{MW}}$) and the cavity ($g_{f}$). We assume that $\ket{E}\leftrightarrow\ket{f}$ is a closed transition and for simplicity we also assume that $\ket{E_{2}}\leftrightarrow\ket{g}$ is a closed transition but this is not a necessity. The figure also indicates how the levels could be realized in ${}^{87}$Rb. Here $\ket{r^{(e)},r^{(e)}}$ with $r=1,2,3$ refers to state $\ket{\text{F}^{(e)}=r,\text{m}_{\text{F}}^{(e)}=r}$ in $5^{2}S_{1/2}$ $(5^{2}P_{3/2})$. (b) Effective three level atom realized by mapping the two-photon drive to an effective decay rate $\tilde{\gamma}_{g}$ and an effective drive $\tilde{\Omega}$}
\label{fig:figureS2}
\end{figure}

We now describe the details of the implementation where the auxiliary atoms is driven by a two-photon process as shown in \figref{fig:figureS2} (reproduced from Fig. 4(a) in the article) in order to suppress the dominant undetectable error caused by spontaneous decay of the auxilliary atom into the state $\ket{g}$ ($\hat{L}_{g}$).

The Hamiltonian in a proper rotating frame is
\begin{eqnarray} \label{eq:supphamil2}
\hat{H}&=&\hat{H}_{e}+\hat{V}+\hat{V}^{\dagger}, \\
\hat{H}_{e}&=&\Delta_{E}\ket{E}\bra{E}+\Delta_{E2}\ket{E_{2}}\bra{E_{2}}+g_{f}(\hat{a}\ket{E}\bra{f}+H.c) \nonumber \\
&&+\frac{\Omega_{\text{MW}}}{2}(\ket{E}\bra{E_{2}}+H.c.) \nonumber \\
&&+\sum_{k}\Delta_{e}\ket{e}_{k}\bra{e}+g(\hat{a}\ket{e}_{k}\bra{1}+H.c.), \\
\hat{V}&=&\frac{\Omega}{2}\ket{E_{2}}\bra{g},
\end{eqnarray}  
where we have now defined $\Delta_{E}=\omega_{E}-\omega_{g}-\omega_{laser}-\omega_{\text{MW}}$,  $\Delta_{E2}=\omega_{E2}-\omega_{g}-\omega_{laser}$ and $\Delta_{e}=\omega_{e}-\omega_{g}-\omega_{laser}-\omega_{\text{MW}}+\omega_{f}-\omega_{1}$. Here $\omega_{laser}$ is the frequency of the laser drive ($\Omega$), $\omega_{\text{MW}}$ is the frequency of the microwave field ($\Omega_{\text{MW}}$) and otherwise $\omega_{x}$ is the frequency associated with level $x$. We assume that the frequency of the cavity is $\omega_{c}=\omega_{laser}+\omega_{\text{MW}}+\omega_{g}-\omega_{f}$ such that the three photon Raman transition from $\ket{g}\to\ket{f}$ is resonant. We have assumed that $\Delta_{E2}$ is large and positive such that the rotating wave approximation is valid for the microwave field. 
The Lindblad operators describing the system are the same as described below Eq.~\eqref{eq:hamil1} except that $\hat{L}_{g}\to\sqrt{\gamma_{g}}\ket{g}\bra{E_{2}}$. Assuming a weak drive $\Omega$, we can follow the same recipe as before to find the following effective operators describing the dynamics of the system. 

\begin{eqnarray}
\hat{H}^{(2)}_{\text{eff}}&=&\sum_{n=0}^{N}\frac{-\Omega^{2}}{4\gamma}\mathrm{Re}\left\{\frac{\tilde{\Delta}_{e}(i\tilde{\Delta}_{E}/2+C_{f})+n\tilde{\Delta}_{E}C}{\tilde{\Delta}_{E2}\tilde{\Delta}_{e}(i\tilde{\Delta}_{E}/2+C_{f})+n\tilde{\Delta}_{E}\tilde{\Delta}_{E2}C-i\tilde{\Delta}_{e}\tilde{\Omega}_{\text{MW}}^{2}/8-n\tilde{\Omega}_{\text{MW}}^{2}C/4}\right\}\ket{g}\bra{g} \otimes \hat{P}_{n} \nonumber \\
&=&\sum_{n=0}^{N}\Delta^{(2)}_{n}\ket{g}\bra{g} \otimes \hat{P}_{n} \label{eq:supham3}\\
\hat{L}_{0}^{\text{eff}(2)}&=&\sum_{n=0}^{N}\frac{-1}{4\sqrt{\gamma}}\frac{\sqrt{C_{f}}\tilde{\Delta}_{e}\Omega\tilde{\Omega}_{\text{MW}}}{\tilde{\Delta}_{E2}\tilde{\Delta}_{e}(i\tilde{\Delta}_{E}/2+C_{f})+n\tilde{\Delta}_{E2}\tilde{\Delta}_{E}C-i\tilde{\Delta}_{e}\tilde{\Omega}_{\text{MW}}^{2}/8-n\tilde{\Omega}_{\text{MW}}^{2}C/4}\ket{f}\bra{g}\otimes \hat{P}_{n} \nonumber \\
&=&\sum_{n=0}^{N}r_{0,n}^{\text{eff}(2)}\ket{f}\bra{g}\otimes \hat{P}_{n} \label{eq:subleff3} \\
\hat{L}_{g}^{\text{eff}(2)}&=&\sum_{n=0}^{N}\frac{\Omega}{2}\frac{\tilde{\Delta}_{e}(i\tilde{\Delta}_{E}/2+C_{f})+n\tilde{\Delta}_{E}C}{\tilde{\Delta}_{E2}\tilde{\Delta}_{e}(i\tilde{\Delta}_{E}/2+C_{f})+n\tilde{\Delta}_{E2}\tilde{\Delta}_{E}C-i\tilde{\Delta}_{e}\tilde{\Omega}_{\text{MW}}^{2}/8-n\tilde{\Omega}_{\text{MW}}^{2}C/4}\frac{\sqrt{\gamma_{g}}}{\gamma}\ket{g}\bra{g}\otimes \hat{P}_{n} \nonumber \\
&=&\sum_{n=0}^{N}r_{g,n}^{\text{eff}(2)}\ket{g}\bra{g}\otimes \hat{P}_{n} \\
\hat{L}_{f}^{\text{eff}(2)}&=&\sum_{n=0}^{N}-\frac{\Omega}{4}\frac{(i\tilde{\Delta}_{e}/2+nC)\tilde{\Omega}_{\text{MW}}}{\tilde{\Delta}_{E2}\tilde{\Delta}_{e}(i\tilde{\Delta}_{E}/2+C_{f})+n\tilde{\Delta}_{E2}\tilde{\Delta}_{E}C-i\tilde{\Delta}_{e}\tilde{\Omega}_{\text{MW}}^{2}/8-n\tilde{\Omega}_{\text{MW}}^{2}C/4}\frac{\sqrt{\gamma_{f}}}{\gamma}\ket{f}\bra{g} \otimes \hat{P}_{n} \nonumber \\
&=&\sum_{n=0}^{N}r_{f,n}^{\text{eff}(2)}\ket{f}\bra{g} \otimes \hat{P}_{n}\\
\hat{L}_{k}^{\text{eff}(2)}&=&\sum_{n=0}^{N-1}\frac{1}{4\sqrt{\gamma}}\frac{\sqrt{C_{f}}\sqrt{C}\tilde{\Omega}_{\text{MW}}\Omega}{\tilde{\Delta}_{E2}\tilde{\Delta}_{e}(i\tilde{\Delta}_{E}/2+C_{f})+n\tilde{\Delta}_{E2}\tilde{\Delta}_{E}C-i\tilde{\Delta}_{e}\tilde{\Omega}_{\text{MW}}^{2}/8-n\tilde{\Omega}_{\text{MW}}^{2}C/4}\ket{f}\bra{g}\otimes\ket{\tilde{o}}_{k}\bra{1}\otimes \hat{P}_{n} \nonumber \\
&=&\sum_{n=0}^{N-1}r_{n}^{\text{eff}(2)}\ket{f}\bra{g}\otimes\ket{\tilde{o}}_{k}\bra{1}\otimes \hat{P}_{n}, \label{eq:subleff4} 
\end{eqnarray}
where we have defined the complex detuning $\tilde{\Delta}_{E2}\gamma=\Delta_{E2}-i\gamma_{g}/2$ and the parameters $r_{0,n}^{\text{eff}(2)},r_{g,n}^{\text{eff}(2)},r_{f,n}^{\text{eff}(2)}$ and $r_{n}^{\text{eff}(2)}$ to characterize the decay described by the Lindblad operators.

We are interested in the limit of large detuning $\Delta_{E2}$ and large cooperativity $C$. In this limit, we find that the dynamics of the system can be mapped to a simple three level atom with effective driving $\tilde{\Omega}\sim\Omega\Omega_{\text{MW}}/(2\Delta_{E2})$ and an effective decay $\tilde{\gamma}_{g}\sim\gamma_{g}\Omega_{\text{MW}}^{2}/\Delta_{E2}^{2}$ as shown in \figref{fig:figureS2}. In principle, the effective operator $\hat{L}_{0}^{\text{eff}(2)}$ leads to an effective decay rate of $\tilde{\gamma}=\gamma_{g}\Omega^{2}/\Delta_{E2}^{2}$ to lowest order in $C$  but we find that this first order term do not destroy the coherence between the qubit states since it is independent of $n$. There is, therefore, no effect of these scattering events and the performance of the gate behaves as if there is an effective decay rate of $\tilde{\gamma}_{g}\sim\gamma_{g}\Omega_{\text{MW}}^{2}/\Delta_{E2}^{2}$. Note that we also have an AC stark shift imposed on the level $\ket{g}$ by the laser characterized by $\Omega$. This will give an overall phase to the system, $\sim\Omega^{2}/(4\Delta_{E2})t$, which we can neglect since it does not influence the gates. 
Since we can do the mapping to the simple three level atom, we find similar results for the performance of the gates for the two-photon scheme as for the simple three level scheme only with effective decay $\tilde{\gamma}_{g}$ and drive $\tilde{\Omega}$ given by the two-photon process. Note, however, that we now assume $\gamma_{g}>0$, which introduces an undetectable error as previously mentioned. We find that this introduces an error in the fidelity of both gates of roughly 
\begin{eqnarray}
&\sim&\frac{(\alpha^{2}-4\alpha\beta-6\beta^{2})\pi^{2}}{128\beta^{2}}\frac{\gamma_{g}^{4}}{\gamma^{4}\Delta_{E2}^{4}}+\frac{(\alpha^{2}+4\alpha\beta+6\beta^{2})\pi}{16\sqrt{\alpha\beta}(\alpha+2\beta)(\alpha+5\beta))}\frac{\gamma_{g}\Omega_{\text{MW}}^{2}}{\gamma\Delta_{E2}^{2}}\frac{1}{\sqrt{C}}.
\end{eqnarray}  
Nonetheless, this error can be suppressed arbitrarily much be increasing $\Delta_{E2}$, which enable us to have a heralded CZ-gate with arbitrarily small error in a realistic atomic setup using the two-photon drive. 

\section{Gate time}
Here we address the question of how strongly we can drive the system and still maintain the validity of perturbation theory. We need to adress this question since the gate time depends inversely on the driving strength as shown in the article and hence this limits the achievable gate time. A necessary criterion for our pertubation theory to be valid is that the energy shifts $\Delta_{n}$ (see Eq.~\eqref{eq:supham2} and Eq.~\eqref{eq:supham3}) are small compared to the driving, i.e. $\sim\Delta_{n}^{2}/\Omega^{2}\ll1$. From Eq.~\eqref{eq:supham2} we find that $\Delta_{n}^{2}/\Omega^{2}\sim\Omega^{2}/(16\Delta_{E}^{2})$ to leading order in the cooperativity $C$ and this criterion is therefore met for $\Omega\ll4\Delta_{E}$. Similarly, for the two-photon process, we find from Eq.~\eqref{eq:supham3} that $(\Delta_{n}^{(2)})^{2}/\Omega^{2}\sim\Omega^{2}/(16\Delta_{E2}^{2})$, to leading order in $C$. Here we thus need $\Omega\ll4\Delta_{E2}$. 

Another criterion need to be meet in order for our perturbative theory to be valid. If none of the qubits couple, we are effectively driving the auxiliary atom/cavity system into a dark state of the form $\cos(\theta)\ket{0,g}-\sin(\theta)\ket{1,f}$, where the mixing angle is $\theta\sim\Omega/g$. Here the number refers to the number of cavity photons. To adiabatically eliminate the state $\ket{f,1}$ from the Hamiltonian as we have done in the effective Hamiltonian requires that $\Omega/g\ll1$. Mapping this criterion to the effective three level scheme realized in the two-photon scheme gives $\Omega\Omega_{\text{MW}}/(2\Delta_{E2}g)\ll1$. 

Finally, we need to consider the scattering of photons from the level $\ket{E2}$ in the two-photon scheme. If the number of scattering events, $n_{scat}$ is large compared to $\Omega^{2}/(16\Delta_{E2}^{2})$, the perturbation theory is not valid even though the other criterions are met. We find that $n_{scat}\sim\frac{12\sqrt{C}\gamma^{2}}{\Omega_{\text{MW}}^{2}}$ for the CZ-gate and we thus need to have  $\frac{3\sqrt{C}\gamma^{2}\Omega^{2}}{\Delta_{E2}^{2}\Omega_{\text{MW}}^{2}}\ll1$

The different criterions for the validity of the perturbation theory are summarized in \tabref{tab:tableS1}. 
\begin{table} [H]
\centering
\begin{tabular}{|c|c|}
\hline
Simple scheme & Two-photon scheme  \\ \hline
$\Omega/(4\Delta_{E})\ll1$ & $\Omega/(4\Delta_{E2})\ll1$ \\ \hline
$\Omega/g\ll1$ & $\Omega\Omega_{\text{MW}}/(\Delta_{E2}g)\ll1$ \\ \hline
- & $3\sqrt{C}\gamma^{2}\Omega^{2}/(\Delta_{E2}^{2}\Omega_{\text{MW}}^{2})\ll1$  \\ \hline
\end{tabular}
\caption{The criterions for our perturbation theory to be valid. }
\label{tab:tableS1}
\end{table}
For all the gate schemes, we have assumed that $\Delta_{E}\propto\sqrt{C}\gamma$. The first criterion for the simple scheme (see \tabref{tab:tableS1}) can thus be met with a driving of $\Omega=a\gamma\sqrt{C}$ where $a/4\ll1$. This driving results in a gate time that decreases as $1/\sqrt{C}$. The value of $a$ will determine the size of the non-adiabatic error. The second criterion is also fulfilled for this driving as long as $\sqrt{\gamma/\kappa}\ll1$. In realistic systems such as the nanocavity system described in Ref.~\cite{thompson, Tiecke}, the ratio $\kappa/\gamma$ can be on the order of 100-1000.    
For the two-photon scheme, we find from \tabref{tab:tableS1} that we can choose $\Omega=a_{2}\Delta_{E2}, \Omega_{\text{MW}}=b\gamma C^{1/4}$ where the constants $a_{2},b$ will determine the size of the non-adiabatic errors as before. Similar to the situation in the simple scheme we assume that $\sqrt{\gamma/\kappa}\ll1$.

\section{Numerical simulation}
In order to confirm our results, we numerically integrated the full Master
equation, defined by the Hamiltonian $\hat H$ and the Lindblad operators,
$\hat L_j \in \{\hat L_0, \hat L_g, \hat L_f, \hat L_1, \hat L_2\}$, 
\bel
\label{eq:full Master eq}
	\frac{d}{dt}\rho(t) = -\frac{i}{\hbar}[\hat H,\rho(t)]+\sum_j \frac{1}{2}
	\left[2 \hat L_j \rho(t) \hat L_j^{\dag} - \rho(t) \hat L_j^{\dag} \hat L_j -
	\hat L_j^{\dag} \hat L_j \rho(t)\right]
\eel
We used the QuTiP 2
package \cite{Johansson20131234}, for Python, to set up the problem and used its
12th-order numerical integration algorithm to find the solution $\rho(t)$ as a
time series.
Then, we used the routines of the same package to analyze the results (see Ref. \cite{peter}). 

For each time series $\rho(t)$, we determined the gate time $t_\text{gate}$, the
success probability $P_\text{success}$, and the fidelity $F$.
We picked $\ket{\psi_0}_{12} =
\frac{1}{\sqrt{2}}\big(\ket{0} +
\ket{1}\big)_1\otimes\frac{1}{\sqrt{2}}\big(\ket{0}+ \ket{1}\big)_2$ as the
initial state of the two qubits, $\ket{g}$ for the control atom, and zero
photons in the cavity. Starting from here, we let the system evolve under the
Master equation Eq.~(\ref{eq:full Master eq}), and determined $P_g(t)$, the
conditional state $\rho_g(t)$ and $F(t)$ as a function of time:
\ba
	P_g(t) &=& \text{Tr}	\Big[\rho(t)\ket{g}\bra{g}\Big],
	\\
	\rho_g(t) &=& \frac{\ket{g}\bra{g}\rho(t)
	\ket{g}\bra{g}}{P_g(t)},
	\\
	F(t) &=& \max_{\phi_1, \phi_2}\Ev{\left.\psi_t^{\phi_1, \phi_2}\right|
	\text{Tr}_\text{c, c}\big(\rho_g(t)\big) \left| \psi_t^{\phi_1, \phi_2}\right.}
\ea
where $\text{Tr}_\text{c, c}$ is the partial trace operation over
the control atom and the cavity, and $\ket{\psi_t^{\phi_1, \phi2}}$ is the
target state transformed with two single qubit $z$-rotations:
\be
	\Ket{\psi_t^{\phi_1, \phi_2}} = \hat U_1(\phi_1) \hat U_2(\phi_2)
	\frac{1}{2}\Big(\ket{00} + \ket{01} + \ket{10} - \ket{11}\Big),
\ee
where $\hat U_{k}(\phi_k) = \exp\left[i \ket{1}_k\bra{1}_k \phi_k\right]$ is the
$z$-rotation of qubit $k$ ($= 1,2$) by the angle $\phi_k$.
From these time series, we determined the gate time $t_\text{gate}$ by
finding the timepoint where $F(t)$ is maximal,
\be
	t_\text{gate} = \underset{t}{\text{argmax}}\;F(t)
\ee
 The fidelity and the success
probability of the gate is then defined as $F = F(t_\text{gate})$,
$P_\text{success} = P_g(t_\text{gate})$. 

Plots of Fig.~\ref{fig:t_gate,P_success} show the gate time ($t_\text{gate}$) and
the success probability ($P_\text{success}$) as a function of $a$, for $\gamma_g
= 0$, $\gamma = 0.01\kappa$, $\Omega=a\gamma\sqrt{C}$ with $C\in\{10,30,100,300,1000\}$. The
detunings, $\Delta_E$ and $\Delta_e$ were chosen to be close to their optimal
value, determined from the adiabatic theory, and numerically optimized to result
in identical effective $\ket{g}\rightarrow \ket{f}$ transition rates  $\Gamma_0
= \Gamma_1 = \Gamma_2$ for the qubit sectors $\ket{00}, \ket{01}, \ket{11}$. The
rates $\Gamma_j$ were found by numerically diagonalizing the master equation for
the qubit sectors separately, and finding the eigenvalue with the smallest (but
non-zero) absolute real part. This numerical optimization yielded the maximal
fidelity. The symbols correspond to the numerical result, whereas the solid
lines show the theoretical values.
The agreement of the results confirms the validity of the adiabatic theory for a driving $\Omega=a\gamma\sqrt{C}$ for $C\lesssim 1000$ and $a\lesssim0.25$. Note, however, that Fig.~\ref{fig:t_gate,P_success} shows how the success probability deviates from the adiabatic result for $a\gtrsim0.25$. A weak increase of this deviation with the cooperativity is seen but from simulations at high C, we believe that this can be removed by gradually ramping $\Omega$ up and down to maintain adiabaticity at the beginning and end of the driving pulse. This was not included in the simulations behind Fig.~\ref{fig:t_gate,P_success} for simplicity.     
 \begin{figure}[h]
\centering
	\includegraphics[width=0.48\textwidth,height=2.9in]{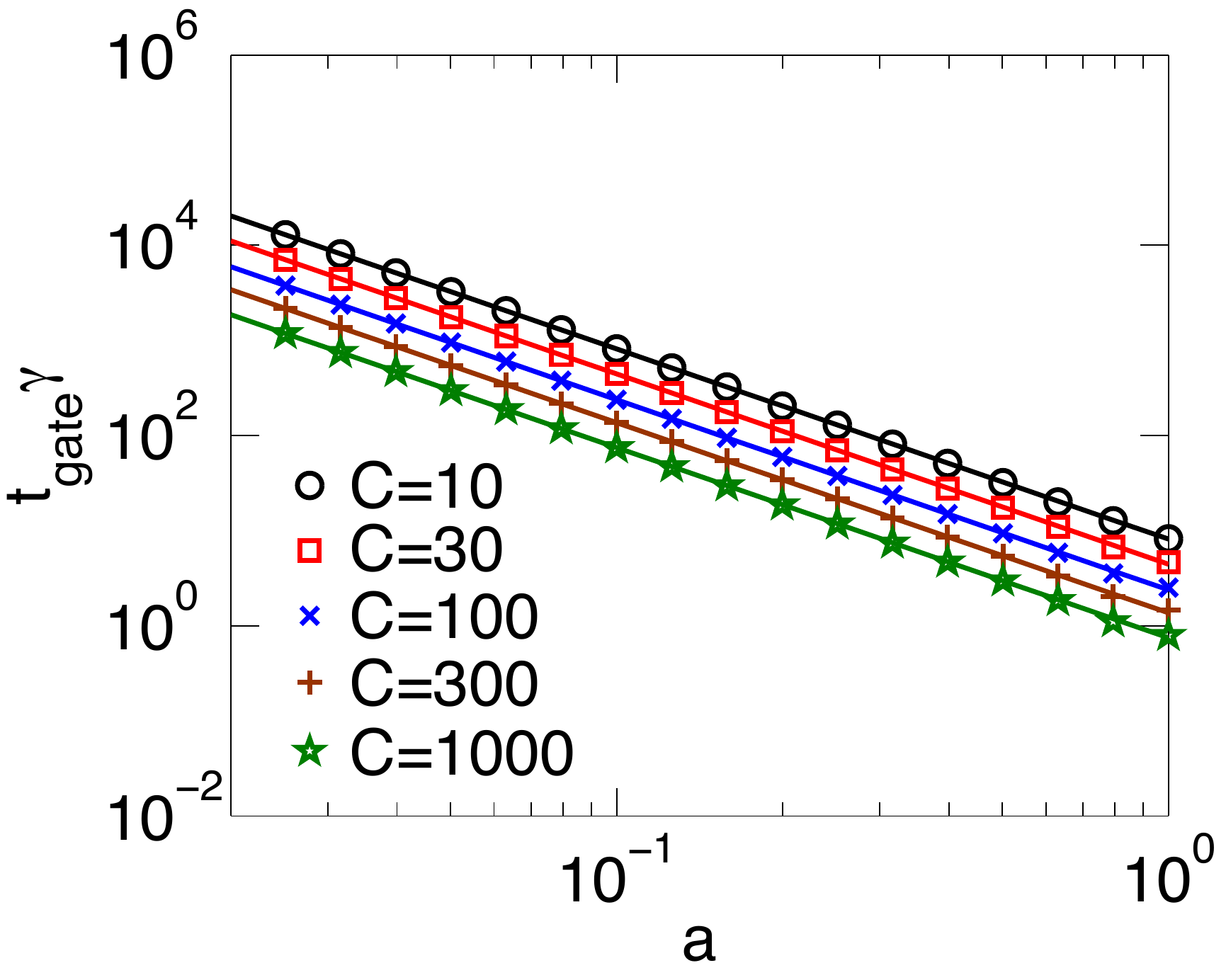}\quad 
	\includegraphics[width=0.48\textwidth,height=2.85in]{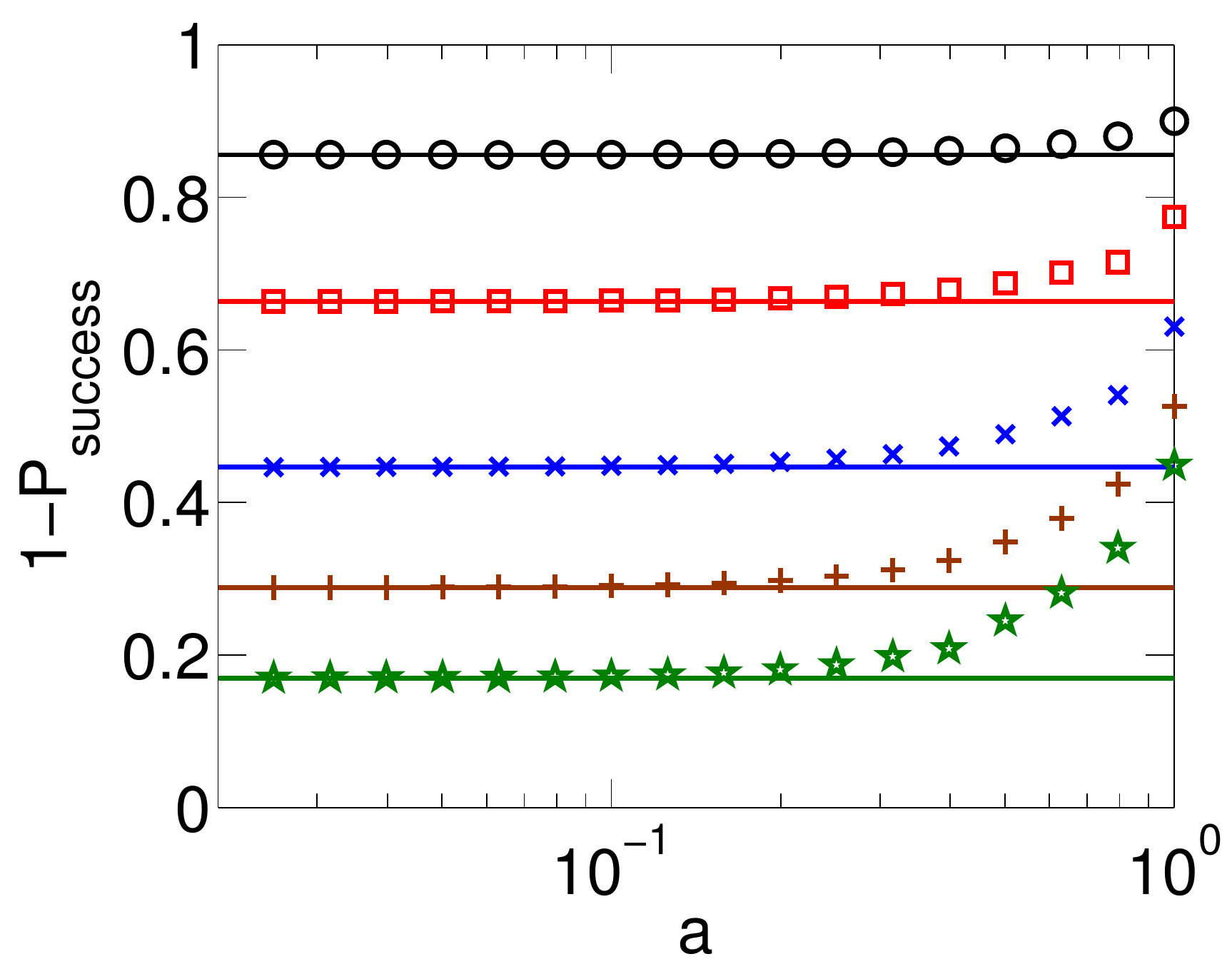}
	\caption{
	\label{fig:t_gate,P_success}
	Gate time (left) and failure probability (right) as a function of
	driving strength ($a$) for $\gamma/\kappa = 0.01$, $\gamma_g =
	0$, and $C\in\{10,30,100,300,1000\}$. The driving strength was assumed to be $\Omega=a\gamma\sqrt{C}$.} 
\end{figure} 
 
\begin{figure}[h] 
\centering
	\includegraphics[width=0.5\textwidth]{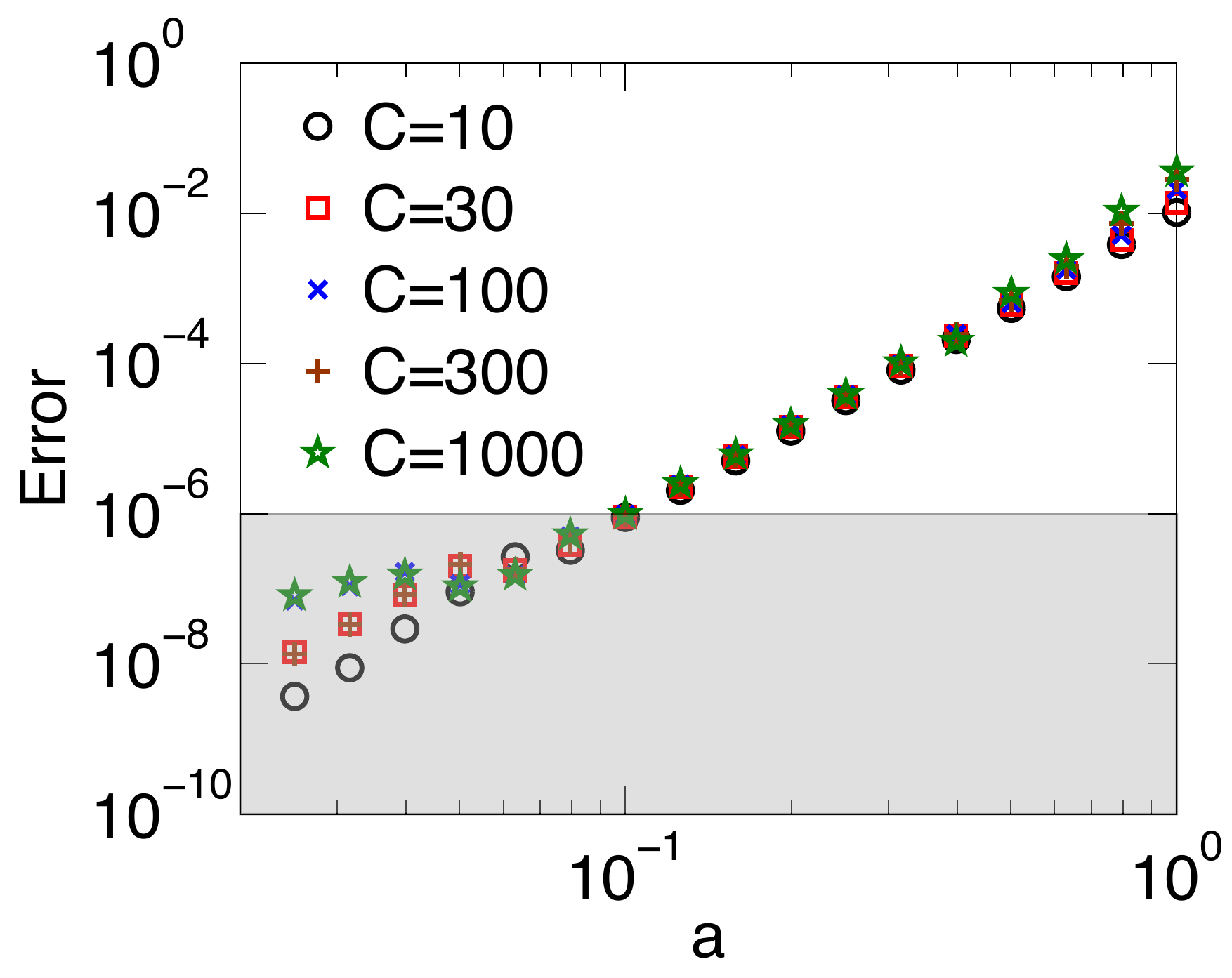}\quad 
	\caption{ 
	\label{fig:F}
	Conditional infidelity of the gate as a function of
	the driving strength $(a)$ for $\gamma/\kappa = 0.01$, $\gamma_g =
	0$, and $C\in\{10,30,100,300,1000\}$. The shaded region (at 
	$\sim 10^{-6}$) shows the limit of numerical accuracy. The driving strength was assumed to be $\Omega=a\gamma\sqrt{C}$.}
\end{figure}

Fig.~\ref{fig:F} shows the conditional \emph{in}fidelity of the gate as a
function of $a$ for the same parameters. The simulation confirms that using $a = 0.25$ is enough to push the (conditional)
infidelity of the gate below $4\cdot10^{-5}$. The fidelity is limited by non-adiabatic effects, which can be suppressed by decreasing $\Omega$ as shown in the figure. Adiabatically ramping Ω up and down at the beginning and end of the gate will also improve the adiabaticity but, for simplicity, we have not included this in the simulations described here. Note, however that for high $C$ ($C>1000$), we find that this gradual ramping of $\Omega$ significantly decreases the non-adiabatic error. 

We repeated the above analysis for the two-photon-driving Hamiltonian in Eq.~\eqref{eq:supphamil2}. We chose 
$ \gamma = \gamma_g = \gamma_f = 0.01\,\kappa$,  $\Omega = \frac{\Delta_{E2}}{8
C^{1/4}}$, and $\Omega_{\text{MW}} = 4\gamma C^{1/4}$, 
and chose $\Delta_E$ and $\Delta_e$ detunings again close to their adiabatic
optimum, but numerically optimized them with the same procedure as previously.
Plots of Fig.\ref{fig:t,P 2} show the gate time and the success probability as a
function of $\Delta_{E2}$ for $C\in\{10, 20,
50,100\}$. Symbols indicate the numerical results while solid lines show the
theoretical values.
 \begin{figure}[h]
\centering
	\includegraphics[width=0.48\textwidth,height=2.85in]{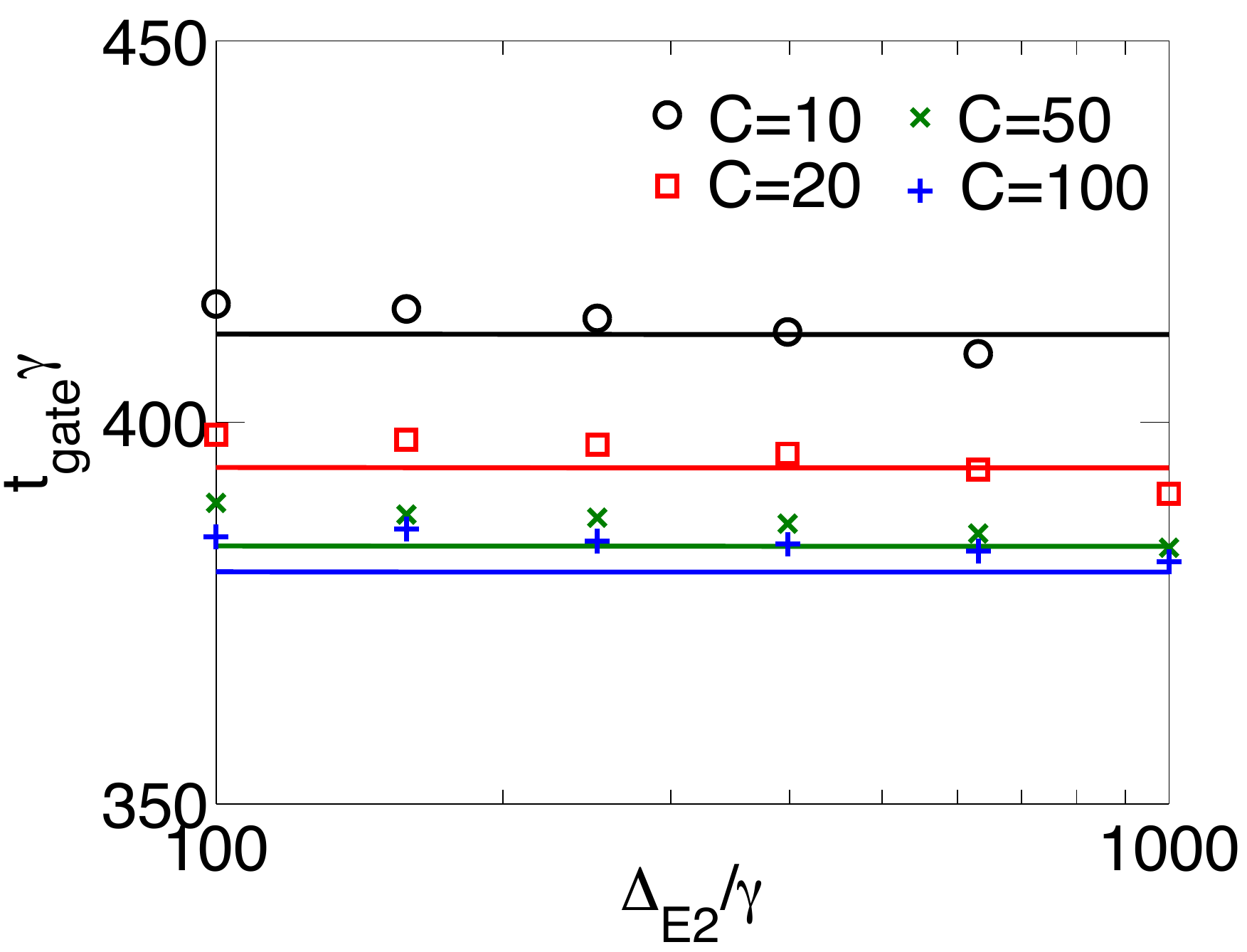}\quad 
	\includegraphics[width=0.48\textwidth,height=2.80in]{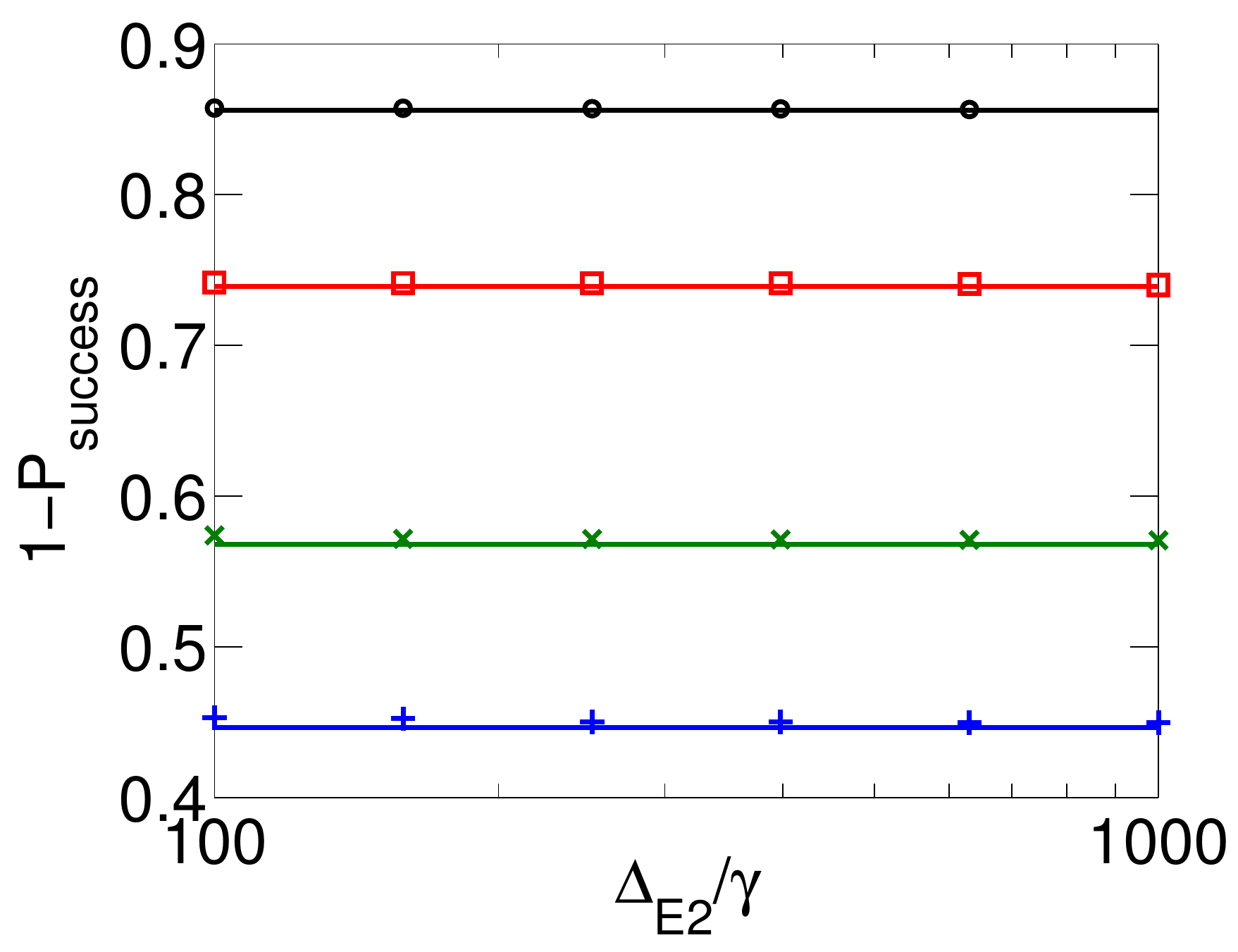}
	\caption{
	\label{fig:t,P 2} 
	Gate time (left) and failure probability (right) as a function of $\Delta_{E2}$ for $\gamma = \gamma_g = \gamma_f = 0.01\kappa$, $\Omega =
	\frac{\Delta_{E2}}{8 C^{1/4}}$, $\Omega_{\text{MW}} = 4\gamma C^{1/4}$, 
	$C = \{10,20,50,100\}$.
	}
\end{figure} 
Fig.~2b in the article shows the conditional \emph{in}fidelity of the
two-photon-driven gate as a function of $C$ for the same choice of parameters.
With these results, we confirm that by increasing $\Delta_{E2}$ we can lower the infidelity error to an arbitrary small level. The gate time is a constant of the cooperativity since we have increased $\Omega_{MW}$ as $C^{1/4}$ in the simulations. We do see some deviation from the analytical results due to non-adiabatic effects, which could be suppressed by decreasing $\Omega$ at the expense of an increase in the gate time. Finally a gradual ramping of $\Omega$ could also decrease the non-adiabatic errors but for simplicity, we have not included this in the simulations. 

\section{Additional errors} 

There are some additional errors in a realistic atomic setup that we have not treated in detail so far. Here we estimate the dominant errors and determine under which conditions, they can be sufficiently suppressed such that they do not limit the performance of the gates. We assume a realistic atomic setup where ${}^{87}$Rb atoms are used both for the auxiliary atom and the qubit atoms. In the ${}^{87}$Rb atoms, we assume that $\ket{g}=\ket{1,1}, \ket{f}=\ket{2,2}$ and $\ket{E_{2}}=\ket{2^{e},2^{e}}, \ket{E}=\ket{3^{e},3^{e}}$ where $\ket{r^{(e)},r^{(e)}}$ with $r=1,2,3$ refers to state $\ket{\text{F}^{(e)}=r,\text{m}_{\text{F}}^{(e)}=r}$ in $5^{2}S_{1/2}$ $(5^{2}P_{3/2})$. In this case, we estimate that the dominant errors are:
\begin{itemize}
\item In our perturbative theory, we have assumed that the laser field ($\Omega$) only couple $\ket{g}\to\ket{E2}$ in the auxiliary atom. However, for a large detuning $\Delta_{E2}$ it may also couple $\ket{f}\to\ket{E}$, which could lead to an undetectable error where the auxiliary atom is pumped back to $\ket{E2}$ from, which it decays to $\ket{g}$. This error is, however, suppressed by the large frequency separation, $\Delta_{g}$ of $\ket{g}$ and $\ket{f}$, which is $\Delta_{g}\sim1000\gamma$ for ${}^{87}$ Rb. We estimate the error using effective operators to find the decay rate back to $\ket{g}$, assuming that the auxiliary atom starts in $\ket{E2}$ and treating the drive $\Omega$ as a perturbation while neglecting the cavity coupling. This is valid as long as $\Delta_{g}\gg\Delta_{E}$, which is fulfilled for $C\lesssim10000$ since $\Delta_{E}\sim\sqrt{C}$. The error increases with $\Delta_{E2}$ but even for $\Delta_{E2}\approx400\gamma$ we find that for $\Omega_{\text{MW}}=4\gamma C^{1/4}$, $\Omega=\Delta_{E2}/8$ the error is $\lesssim10^{-4}$.     

\item The microwave might also couple the ground states $\ket{0}-\ket{1}$ of the qubit atoms and the ground states $\ket{g}-\ket{f}$ of the auxiliary atom. The coupling of $\ket{0}-\ket{1}$ means that the qubit atoms also couple to the cavity even though they are in state $\ket{0}$. We estimate the error from this to be on the order of $\Omega_{\text{MW}}^{2}/(\Delta_{g}-(\Delta_{E2}-\Delta_{E}+\Delta_{2\to3}))^2$ where $\Delta_{2\to3}$ is the splitting between $\ket{E_{2}}$ and $\ket{E}$. For ${}^{87}$Rb, $\Delta_{2\to3}\approx44\gamma$. Below we argue that we need $\Delta_{E}<0$. Since $\Delta_{E}\approx-\sqrt{C}\gamma$ this error will increase slowly with cooperativity but it is suppressed by $\Delta_{g}$. For $\Omega_{\text{MW}}=4\gamma C^{1/4}$, we find that the error is $\lesssim10^{-4}$ for $C\lesssim1000$ even for $\Delta_{E2}\approx400\gamma$. The errors from the coupling of the states $\ket{g}-\ket{f}$ in the auxiliary atom will likewise be suppressed by the large energy splitting $\Delta_{g}$. These errors can also be further suppressed by decreasing $\Omega_{\text{MW}}$ at the cost of a larger gate time. 
\end{itemize}

The above errors can be highly suppressed using e.g.${}^{88}$Sr, ${}^{138}$Ba${}^{+}$ or ${}^{40}$Ca${}^{+}$ instead of ${}^{87}$Rb. For these atoms, the ground states can be encoded in the $S_{0}$ and $P_{0}$ manifoldes for ${}^{88}$Sr and the $S_{1/2}$ and $D_{3/5}$ manifolds for ${}^{138}$Ba${}^{+}$ and ${}^{40}$Ca${}^{+}$, which have separations at optical frequencies between the stable states. 

A final error that we will consider is that the transition $\ket{E}\leftrightarrow\ket{f}$ will not be completely closed if the cavity is linearly polarized. This will, e.g. be the case for the system in Ref. \cite{thompson}. Such a cavity also couples $\ket{f}$ to the states $\ket{1^{e},1^{e}},\ket{2^{e},1^{e}}$ and $\ket{3^{e},1^{e}}$. From $\ket{1^{e},1^{e}}$ and $\ket{2^{e},1^{e}}$ there might be an undetectable decay back to $\ket{g}$, which will introduce an error $\propto1/\sqrt{C}$ in the gates. The probability of an undetectable decay from these states should be compared to the probability of the detectable decays where the cavity photon is scattered of the qubit atoms instead. For ${}^{87}$Rb, we estimate this error by compairing the strengths of the effective couplings from $\ket{f}$ to $\ket{1^{e},1^{e}}$ and $\ket{2^{e},1^{e}}$ with a subsequent decay to $\ket{g}$ with the strength of the effective coupling from $\ket{1}$ to $\ket{e}$ in the qubit atoms with a subsequent decay back to $\ket{1}$. The latter process has a detuning of $\Delta_{E}$ while the first two are additional detuned by the energy gaps between $\ket{3^{e},3^{e}}$ and $\ket{1^{e},1^{e}}$ and $\ket{3^{e},3^{e}}$ and $\ket{2^{e},1^{e}}$ respectively, assuming that $\Delta_{E}<0$. We find that since $\abs{\Delta_{E}}$ grows as $\sqrt{C}$ the error increases from $\sim5\cdot10^{-5}$ at $C=1$  to a maximum value of $\sim2\cdot10^{-3}$ for $C\sim3000$ for which $\Delta_{E}$ is comparable to the extra detunings of the $\ket{1^{e},1^{e}}$ and $\ket{2^{e},1^{e}}$ transitions compared to the $\ket{e}$ transition. For $C>3000$ the error decreases as $1/\sqrt{C}$.
Note that this error could be removed  by making a 4 photon drive from $\ket{g}$ to $\ket{E}$ by letting $\ket{g}=\ket{1,-1}$. Another approach is to consider other atoms such as ${}^{40}$Ca${}^{+}$, with more favorable levelstructures. The state $\ket{g}$ could be encoded in the $3^{2}D_{5/2}$ subspace while the state $\ket{f}$ could be encoded in the $4^{2}S_{1/2}$ subspace and similarly for the qubit states $\ket{0}$ and $\ket{1}$. In such a setup, we will have separations of optical frequencies between the qubit states and we can remove the decay from the excited state back to $\ket{g}$ by, e.g. driving from $3^{2}D_{5/2}$ to $4^{2}P_{1/2}$ through $3^{2}D_{3/2}$. 

\end{document}